\documentclass[referee,useAMS,usenatbib,fleqn]{mnras}
\usepackage{amsopn, natbib} \usepackage[utf8]{inputenc}
\usepackage{amsmath} \usepackage{caption} \usepackage{amsfonts}
\usepackage{amssymb} \usepackage{graphicx} \usepackage{hyperref}
\usepackage{subfigure} \usepackage{psfrag}

\def\vis{\mathcal{V}\left( \mathbfit{U}, \nu \right)}
\def\visnonu{\mathcal{V}\left( \mathbfit{U}\right)}
\def\visnonup{\mathcal{V}^{'}\left( \mathbfit{U}\right)}
\def\v2{V_2\left(\mathbfit{U},\Delta \nu = 0\right)}
\def\viscor{V_2\left( \mathbfit{U}_a,\nu_i; \mathbfit{U}_b,\nu_j\right)}
\def\viscornonu{V_2\left( \mathbfit{U}_a; \mathbfit{U}_b\right)}

\def\viscora{V_2\left( \mathbfit{U}_a,\nu_i; \mathbfit{U}_a,\nu_j\right)}

\def\maps{\mathcal{C}_{\ell}\left( \nu_i, \nu_j \right)}

\def\dBdTnu{Q_{\nu}}
\def\dBdTnonu{Q}
\def\dBdT{\left(\frac{\partial B}{\partial T}\right)}
\def\Blm{B_{\ell}^m\left(\mathbfit{U}, \nu \right)}

\def\Blmnonu{B_{\ell}^m\left(\mathbfit{U} \right)}
\def\BlmRnonu{B_{\ell}^m\left(\U \right)}
\def\Blma{B_{\ell}^m(\U_a )}
\def\alm{a_{\ell}^m}

\def\Ylm{Y_{\ell}^m(\hat{\mathbfit{n}})}

\def\pbnonu{A\left(\Delta \hat{\mathbfit{n}}\right)}

\def\apnonu{\tilde{a}(\U - \U^{'})}
\def\Flm{F_{\ell}^m}
\def\Wl{\mathcal{W}_{\ell}\left(\U_a, \nu_i ;\U_b, \nu_j \right)}
\def\Wlsame{\mathcal{W}_{\ell}\left(\U_a, \nu_i ;\U_a, \nu_j \right)}

\def\Wlsamec{\mathcal{W}_{\ell}\left(\U_a, \nu_c ;\U_a, \nu_c \right)}
\def\Wladjc{\mathcal{W}_{\ell}\left(\U_a, \nu_c ;\U_{a \pm 1}, \nu_c \right)}
\def\U{\mathbfit{U}} 
\newcommand{\HI}{{H{\sc i}~}}
\def\n{\hat{\mathbfit{n}}} 
\def\m{\hat{\mathbfit{p}}}
\def\dn{\Delta \mathbfit{n}}

\def\SHE{SH}

\title{A Spherical Harmonic Analysis of  the Ooty Wide Field Array (OWFA) Visibility Signal}

\author[Chatterjee et al.]{Suman Chatterjee$^{1,2}$\thanks{E-mail: suman05@phy.iitkgp.ernet.in}, 
Somnath Bharadwaj$^{1,2}$\thanks{E-mail: somnath@phy.iitkgp.ernet.in }\\ $^{1}$Department of Physics,
  Indian Institute of Technology Kharagpur, Kharagpur - 721 302,
  India.\\ $^{2}$Centre for Theoretical Studies, Indian Institute of
  Technology Kharagpur, Kharagpur - 721 302, India.}

\begin{document}
\date{\today}

\pagerange{\pageref{firstpage}--\pageref{lastpage}} \pubyear{2018}

\maketitle
\label{firstpage}
\begin{abstract}
Considering redshifted $21$-cm intensity mapping with the upcoming
OWFA whose field of view subtends $\sim 57^{\circ}$ in the N-S
direction, we present   a  formalism which  relates the
measured visibilities to the spherical harmonic coefficients of the
sky signal. We use this to calculate  window functions 
which relate the two-visibility correlations {\it i.e.} the correlation between 
the visibilities measured at two baselines and two frequencies, to different multipoles
of   the multi-frequency angular power spectrum
$C_{\ell}(\nu_1,\nu_2)$.  The formalism here is  validated using 
simulations.    We also  present approximate 
closed form analytical  expressions which can be used to calculate the 
window functions. Comparing the widely adopted
flat sky approximation,  we find that its predictions match those  
of our spherical harmonic formalism to within $16 \%$
across the entire OWFA baseline range. The match improves at large
baselines where we have  $< 5 \%$ deviations.
\end{abstract}

\newpage
\section{Introduction} \label{sec:intro}
Intensity mapping with neutral Hydrogen \HI 21-cm radiation is a
promising tool to study the large scale structures in the
post-reionization Universe \citep{Bharadwaj2001b}. The redshifted \HI 21-cm
observations hold the potential of measuring the Baryon Acoustic
Oscillation (BAO) that is embedded in the power spectrum of \HI 21-cm
intensity fluctuations at all redshifts and the comoving scale of BAO
can be used as a standard ruler to constrain the evolution of the
equation of state for dark energy \citep{Wyithe2008, Chang2008,
  Seo2010, Masui2010}. Further, a measurement of just the \HI 21-cm power
spectrum can also be used to constrain the cosmological parameters
\citep{Bharadwaj2009,Visbal2009}. The higher order statistics such as
the \HI 21-cm bispectrum holds the prospect of quantifying the
non-Gaussianities in the \HI 21-cm signal \citep{Ali2005, Hazra2012}.  
Using the \HI signal in cross-correlation with the WiggleZ galaxy survey data, 
the Green Bank Telescope (GBT) has made the first detection of the \HI 
signal in emission at $z \approx 0.8$ \citep{Chang2010}.

A number of post-reionization experiments are either being planned or are ongoing at present.
The Giant Meterwave Radio telescope \citep[GMRT;][]{Swarup1991} is 
sensitive to the cosmological \HI signal from a range of redshifts 
in the post-reionization era \citep{Bharadwaj2003, 
Bharadwaj2005}. The upgraded GMRT \citep[uGMRT;][]{Gupta2017} is expected 
to have a larger bandwidth for which the prospects of a detection are 
investigated in Chatterjee et. al. 2018(in preparation). The Canadian 
Hydrogen Intensity Mapping Experiment \citep[CHIME;][]{Bandura2014} and 
the Hydrogen Intensity and Real-time Analysis eXperiment \citep[HIRAX;]
[]{Newburgh2016} aims to measure the BAO in the redshift range 
$0.8-2.5$. Future experiments, like Tianlai \citep{Chen2012,Chen2015} and SKA1-MID 
\citep{Bull2015ApJ} also aim to measure the \HI 21-cm 
signal from the post-reionization era.

 The Ooty Wide Field Array (OWFA; \citealt{Subrahmanya2016a}) is an
 upgraded   version of the Ooty Radio Telescope 
 (ORT, \citealt{Swarup1971}). The upgrade will result in two
 concurrently functioning modes named Phase I (PI) and Phase II
 (PII). The primary science goals of OWFA have
 been outlined in \citet{Subrahmanya2016b}, and the measurement of the
 z = 3.35 post-reionization \HI 21-cm power spectrum is one of its
 major objectives \citep{Ali2014}. It has been predicted
  that a $5\sigma$ detection of the amplitude of
 the \HI 21-cm power spectrum is possible with $\sim 150$ hrs of
 observation \citep{Bharadwaj2015}.  Further, \citet{Sarkar2016b} have
 predicted that a $\sim  5\sigma$ measurement of the binned \HI 21-cm
 power spectrum is  possible in the $k$-range $0.05 \, {\rm Mpc}^{-1} \leq k \leq 0.3 \,
 {\rm Mpc}^{-1}$  with $1,000$ hrs of observation. 
 
The primary observable quantity for radio 
telescopes are the visibilities, which are the correlation of the 
measured voltages at each antenna. It is possible
to directly estimate the redshifted \HI 21-cm power spectrum from the visibilities
measured by a low frequency radio interferometric array
\citep{Bharadwaj2001a,Bharadwaj2005}.  \citet{Ali2014} presents 
theoretical predictions for the two-visibility correlations {\it i.e.} the correlation between 
the visibilities measured at two baselines and two frequencies, expected 
at OWFA considering both the \HI $21$-cm signal  and also various
foreground components. \citet{Chatterjee2017} 
and \citet{Marthi2016a} have used numerical simulations to
respectively predict the $21$-cm signal and the  various 
foreground contributions to the two-visibility correlations expected at 
OWFA. Considering OWFA, \citet{Sarkar2016b} presents an analytic
technique to simulate the expected $21$-cm visibility signal.  
The common assumption of all the earlier works mentioned here 
is that the field of view (FoV) of the telescope is sufficiently
small so that the observed sky can be assumed to be
flat. In fact the flat sky approximation (FSA) is an underlying 
assumption in a large fraction of the works related to 
measuring the cosmological $21$-cm power spectrum
(e.g. \citealt{Morales2004}, \citealt{Ali2015}). 
Considering OWFA (both PI and PII), the FoV covers $1.8^{\circ}$ in
the E-W direction while this is $4.8^{\circ}$ and $28.6^{\circ}$ 
in the N-S direction for PI and PII respectively. We expect the
FSA to be a reasonably good approximation for PI, however for PII the
large N-S extent of the FoV brings to question the validity of this
assumption. It is desirable  to consider the spherical nature of the
sky in making predictions for PII.

\citet{Shaw2014} have introduced the ``m-mode'' formalism which  
incorporates the  spherical nature of the sky. This essentially deals
with drift scan observations, however their results can also be used
to predict the signal for observations where the telescope tracks a
fixed region of the sky. \citet{Liu2016} have introduced a spherical
Fourier-Bessel technique for analysing the 21-cm power spectrum,   
that incorporates the spherical nature of the sky. 
\citet{Zhang2016a} and \citet{Ghosh2017}
present sky map reconstruction methods based on the spherical
harmonics (SH) transformation. 

In this work, we develop a spherical sky formalism which is
particularly suited for telescopes like OWFA, where the baselines are
all coplanar with the antenna aperture. The formalism relates the
visibilities to the SH coefficients of the sky signal
through the ``beam transfer function''. Section~\ref{sec:OWFA} of this
paper presents  a brief overview  of OWFA, whereas the formalism is
presented in Section~\ref{sec:formalism}. In addition to the numerical 
evaluation of the beam transfer function, we also use the Limber
approximation (LA) to obtain  an analytical expression for  the same. 
In Section~\ref{sec:vis_cor} we use the formalism to relate the two 
visibility correlations to the  multi-frequency  angular power
spectrum (MAPS, \citealt{Datta2007}) which quantifies the statistics
of the 21-cm signal. In Section~\ref{sec:simulation} we validate our
formalism using  all-sky simulations. We study  the differences
between the FSA and the formalism of this paper in order to
quantify  how important the spherical nature of the sky is for the
different phases  of OWFA. The Results are presented in 
Section~\ref{sec:results} and we present  Summary and 
Conclusion in Section~\ref{sec:discussion}.


\section{OWFA} \label{sec:OWFA}
\begin{figure*}
\psfrag{Equatorial}{blah} \psfrag{xx}{$x$} \psfrag{yy}{$y$}
\psfrag{zz}{$z$} \psfrag{bb}{$b$} \psfrag{dd}{$d$}
\psfrag{mm}{$\m$} \psfrag{nn}{$\hat{\mathbfit{n}}$}
\psfrag{rrp}{$\mathbfit{r}'$} \centering
\includegraphics[scale=0.4]{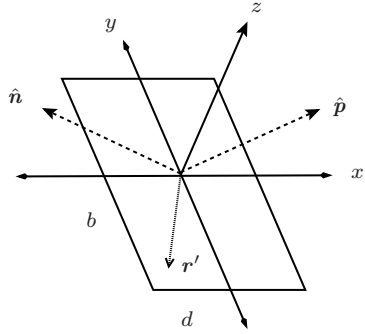}
\caption{This shows the aperture of one of the OWFA antennas. The unit
  vector $\m$ is the pointing direction of the antenna
  beam, $\hat{\mathbfit{n}}$ points towards an arbitrary direction and
  $\mathbfit{r}'$ refers to a displacement on the antenna aperture. The
  $x$ and $y$ axes are along the N-S and E-W directions respectively,
  while the $z$ axis, which is normal to the aperture, points towards
  the celestial Equator ($\delta=0$).}
\label{fig:OWFA_aperture}
\end{figure*}

 The Ooty Radio Telescope (ORT) is a $530 \, {\rm m}$ long (N-S) and $30
 \, {\rm m}$ wide (E-W) offset-parabolic cylinder, operating at a nominal
 frequency of $326.5 \, {\rm MHz}$. The telescope is equatorially
 mounted, i.e. the long axis of the reflecting cylinder is in the
 North-South direction, parallel to the Earth's rotation axis. The
 telescope can be mechanically steered along the East-West direction
 by a single rotation along the axis of the cylinder. The feed
 consists of 1056 North-South dipoles of length $0.5 \lambda$ arranged
 nearly end to end along the focal line of the reflector. The
 telescope has a $530 \, {\rm m} \times 30 \, {\rm m}$ rectangular aperture.
 
 Currently this telescope is being upgraded \citep{Subrahmanya2016a} to operate as an
 interferometer the Ooty Wide Field Array (OWFA). The upgrade will
 result in two concurrent modes namely OWFA PI and PII
 respectively. In OWFA PI, the signals from $N_d=24$ adjacent dipoles
 will be combined to form a single antenna. Each antenna has a
 rectangular aperture of dimension $b \times d$, where $b = 30 \, {\rm
   m}$ and $d = 11.5 \, {\rm m}$ respectively. The array will have $N_A =
 40$ such antennas separated by $d = 11.5 \, {\rm m}$, arranged along the
 North-South axis of the cylinder. In OWFA PII, the signals from $N_d=4$ adjacent dipoles will be
 combined to form a single antenna. Each antenna has a rectangular
 aperture of dimension $b \times d$, where $b = 30 \, {\rm m}$ and $d =
 1.92 \, {\rm m}$ respectively. The array will have $N_A = 264$ such
 antennas separated by $d = 1.92 \, {\rm m}$ arranged along the
 North-South axis of the cylinder.  
 
 Figure~\ref{fig:OWFA_aperture}
 shows a schematic diagram of the aperture for one of the OWFA
 antennas. For both PI and PII this is a rectangle of dimensions $b
 \times d$ with $b = 30 \, {\rm m}$, however as discussed earlier the
 dimension $d$ is different for the two modes. In both the modes, the
 digitised signals from $N_d$ successive dipoles are combined to form
 the antenna beam. This allows the pointing direction
 $\m$ of the antenna beam pattern (Figure
 \ref{fig:OWFA_aperture}) to be steered electronically along the
 North-South direction by introducing phases. The antenna primary beam
 pattern $A\left(\Delta \hat{\mathbfit{n}},\nu \right)$ quantifies how
 the individual antenna responds to the signal from different
 directions $\hat{\mathbfit{n}}$ on the sky (Figure
 \ref{fig:OWFA_aperture}), here $\Delta \n =
 \n -\m$. It is possible to calculate
 $A\left(\Delta \hat{\mathbfit{n}},\nu
 \right)$ \citep[e.g][]{Chengalur2007} using,
\begin{equation}
A\left(\Delta \hat{\mathbfit{n}},\nu \right) = \int d^2 \U^{'} \;
e^{2 \pi i \U^{'}\cdot \Delta \hat{\mathbfit{n}} }\;
\tilde{a}\left(\U^{'}, \nu \right),
\label{eq:ft_aperture}
\end{equation}
where, $\U^{'} = \mathbfit{r}'/\lambda$ refers to displacement
$\mathbfit{r}'$ on the antenna aperture and the aperture power pattern
$\tilde{a}\left(\U^{'}, \nu \right)$ is the auto-convolution of
the electric field pattern at the antenna aperture (Figure
\ref{fig:OWFA_aperture}). It is useful to note that the phase 
factor $e^{2 \pi i \U^{'}\cdot \Delta \hat{\mathbfit{n}} }$ in 
eq.~(\ref{eq:ft_aperture}) is invariant under a mirror reflection 
with respect to the aperture plane. It follows that eq.~(\ref{eq:ft_aperture}) 
predicts a primary beam pattern which is exactly identical in the
upper hemisphere $(UH)$ and lower  
hemisphere $(LH)$ of the sky. The telescope, however,  only responds
to the $UH$ and the $LH$ is not accessible to
the telescope. Here we have exploited 
the fact that eq.~(\ref{eq:ft_aperture}) predicts an
identical beam pattern in both the $UH$ and the $LH$ to simplify the
mathematical analysis in subsequent parts of this paper. 

For the purpose of the present analysis we make the simplifying
assumption that the aperture is uniformly illuminated such that the
electric field is uniform everywhere on the $b\times d$ rectangular
aperture of the OWFA antenna. In this case the OWFA the aperture power
pattern can be expressed as \citep{Ali2014},

\begin{equation}
\tilde{a}\left(\U^{'}, \nu \right) = \frac{\lambda^2 }{bd}
\Lambda \left(\frac{U_x^{'}}{d} \lambda \right) \Lambda
\left(\frac{U_y^{'}}{b} \lambda \right) \, ,
\label{eq:OWFA_ap}
\end{equation}
where $\U^{'} = (U_x^{'},U_y^{'})$, $\Lambda(\xi)$ is the triangular 
function defined as, $\Lambda(\xi) = 1-|\xi|$ for $|\xi|<1$ and
$\Lambda(\xi) = 0$ for $|\xi|\geq 1 $ \citep[see Figure (2a)
  of][]{Ali2014}.

We see that for OWFA, $\tilde{a}\left(\U^{'}, \nu \right)$ peaks
at $\U^{'} = 0$ and falls off as $\U^{'}$ is
increased. Further this has compact support and
$\tilde{a}\left(\U^{'}, \nu \right) = 0$ if $\U^{'}$ exceeds
the aperture dimensions. Note that these properties of
$\tilde{a}\left(\U^{'}, \nu \right)$ are not particular to OWFA
alone. In general, for any antenna, we expect
$\tilde{a}\left(\U^{'}, \nu \right)$ to peak at $\U^{'} =
0$, fall off as $\U^{'}$ is increased and to have compact support
whereby $\tilde{a}\left(\U^{'}, \nu \right) = 0$ if $\U^{'}$
exceeds the aperture dimensions.

\begin{figure*}
\psfrag{Equatorial}{blah} \centering
\begin{minipage}{190mm}
\subfigure[Primary beam pattern for PI]{\label{fig:pbeam-PI}
  \includegraphics[scale=0.35,angle=0]{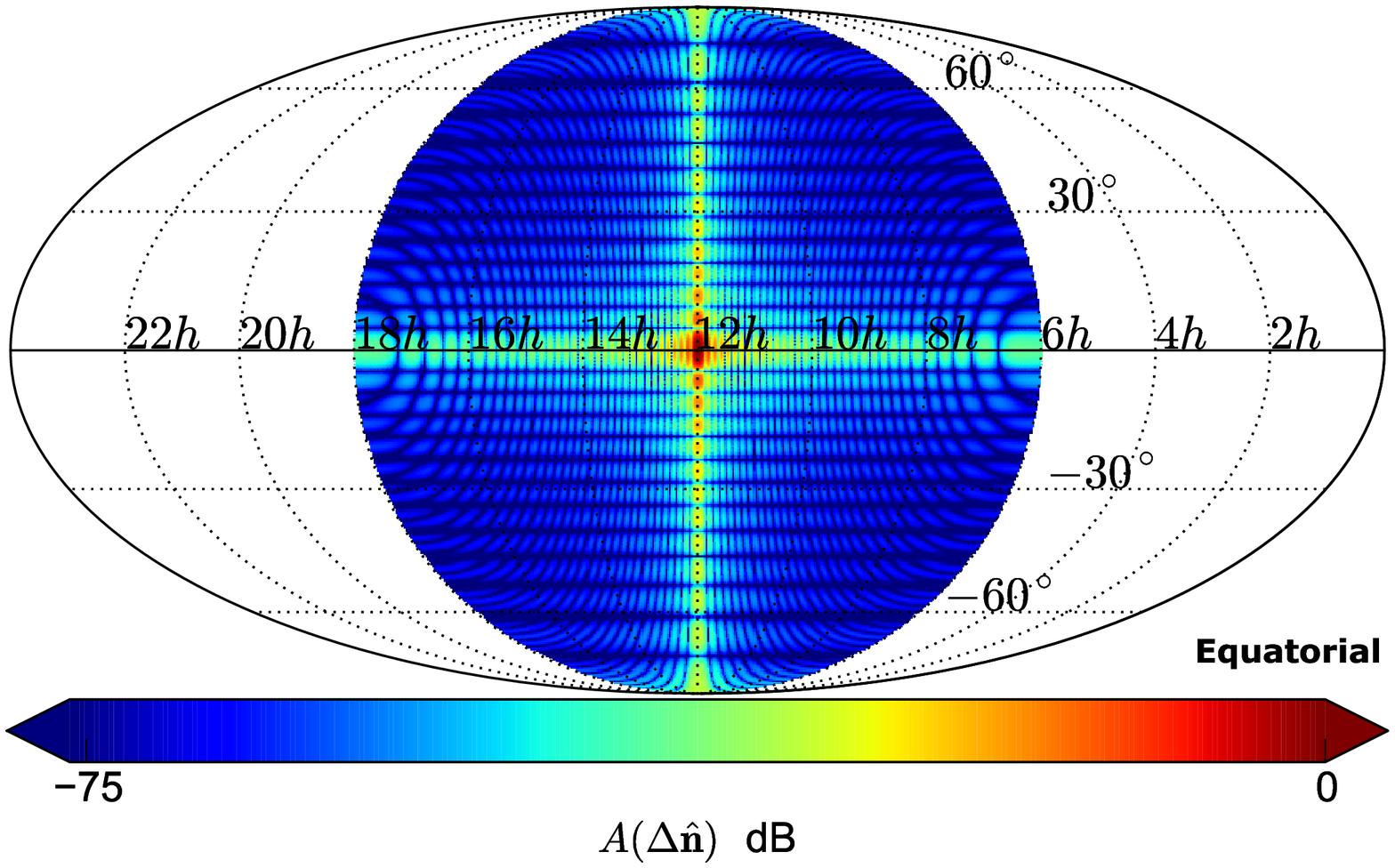}}
\hskip 7mm \subfigure[Primary beam pattern for
  PII]{\label{fig:pbeam-PII}
  \includegraphics[scale=0.35,angle=0]{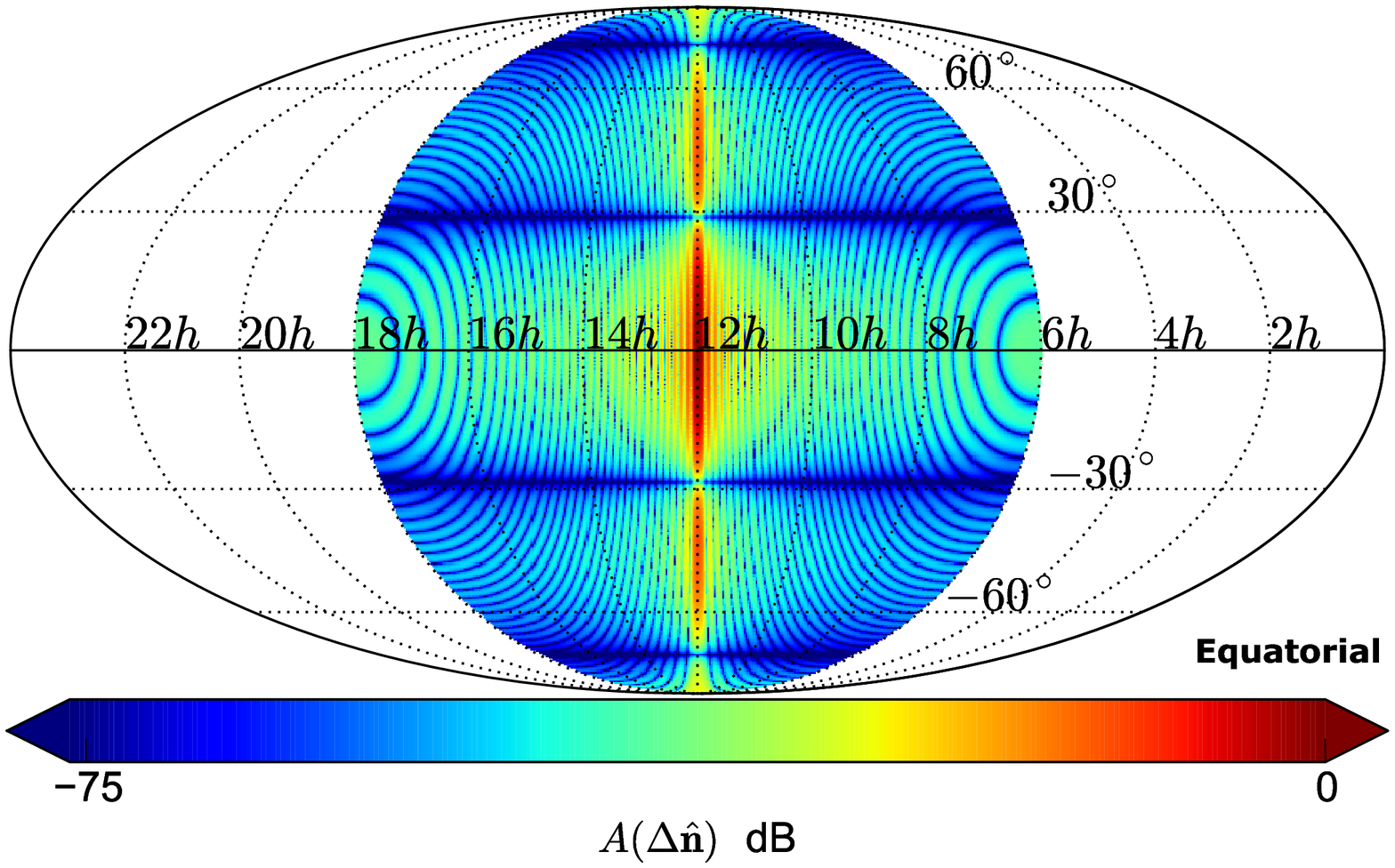}}
\end{minipage}
\caption{The left and right panels  show the  primary beam pattern
  on the visible upper hemisphere for OWFA PI and PII respectively. }
\label{fig:pbeam}
\end{figure*}
Using equation (\ref{eq:ft_aperture}) to calculate the OWFA primary
beam pattern $A(\dn,\nu)$, we get the product of two
sinc-squared functions
\begin{equation}
 A(\dn,\nu)= {\rm sinc}^2 \left( \frac{\pi b \nu \,
   \Delta n_y}{c} \right) {\rm sinc}^2 \left( \frac{\pi d \nu \,
   \Delta n_x}{c} \right)
\label{eq:pb}
\end{equation}
where $\Delta n_y$ and $\Delta n_x$ are respectively the $y$ and $x$
components of $\dn$ \citep{Marthi2016a}.

Figure~\ref{fig:pbeam} shows the primary beam pattern for OWFA PI and
PII as a function of the celestial coordinates $(\alpha,\delta)$ on
the visible upper hemisphere of the sky. Here the telescope beam is assumed
to point towards $\m =(\alpha,\delta) = (0,0)$. The OWFA
beam is the diffraction pattern of a rectangular slit of dimension $b
\times d$. The beam pattern is normalized to $ A(\Delta
\mathbfit{n},\nu)=1$ at the maxima which occurs when $\Delta
\mathbfit{n}=0$ ({\it i.e.} $\mathbfit{n} = \m$), and $ A(\Delta
\mathbfit{n},\nu)$ falls off as $\dn$ is increased. The
angular extent of the main lobe is determined by the first null which
occurs at $\alpha_0 \approx \lambda / b= \pm 1.8^{\circ}$ along
$\alpha$ for both PI and PII, and at $\delta_0 \approx \pm4.8^{\circ}$
and $\pm 28.6^{\circ}$ along $\delta$ for PI and PII respectively.
After the first nulls, there are side lobes in the OWFA primary beam
pattern. The maxima of the $1^{st}$ and $2^{nd}$ side lobes are
respectively $5\%$ and $2\%$ of the maxima of the main lobe.

The Fourier relation (eq.~\ref{eq:ft_aperture}) implies that the
primary beam pattern gets wider if the antenna aperture dimensions are
reduced. Here, we see that for PI the angular extent of the main lobe,
which defines the antenna's FoV, is small and
restricted within a few degrees on the sky. For this small region
($\theta \ll 1$ in radians) it is a reasonably good approximation to
neglect the curvature of the sky, and it is adequate to carry out any
further analysis for OWFA PI using the FSA
\citep[e.g.][]{Marthi2016a}. For PII, however, we find that the FoV
extends across $57.2^{\circ}$ in the N-S direction. It is not possible
to ignore the curvature of the sky over such large angular extent, and
it is necessary to consider the spherical sky for any further analysis
for PII. Considering a radio-interferometer like OWFA, in the next
Section we develop a formalism to analyze the measured visibilities in
terms of a SH  expansion which provide a natural basis for the
signal on a spherical sky.

\section{Visibilities and beam transfer functions} \label{sec:formalism}

\begin{figure*}
\centering \psfrag{nn}{$\hat{\mathbfit{n}}$}
\psfrag{mm}{$\m$} \psfrag{uu}{$\mathbfit{r}$}
\psfrag{dd}{$\mathbfit{r}'$} \psfrag{hh}{$b$} \psfrag{gg}{$d$}
\includegraphics[scale=0.4]{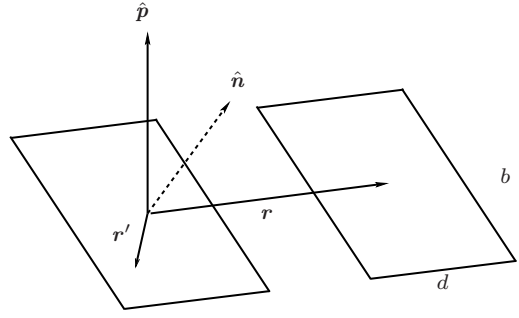}
\caption{This shows a pair of antennas in a radio
  interferometer. Here  $\mathbfit{r}$ is the antenna  separation while 
  $\m$, $\hat{\mathbfit{n}}$ and $\mathbfit{r}'$ are as defined in Figure
\ref{fig:OWFA_aperture}.}
\label{fig:aperture}
\end{figure*}

The complex visibilities $\vis$ are the quantities measured in a radio
interferometric array. Here the baseline $\U$ refers to a pair of
antennas separated by $\mathbfit{r}$ (Figure \ref{fig:aperture}), and
$\U =\mathbfit{r}/\lambda$ is the antenna separation in units of
the observing wavelength. The baselines available at OWFA are
\begin{equation}
\U_n = n \frac{d}{\lambda} \hat{\mathbfit{i}} \quad \quad \left(
1 \leq n \leq N_A - 1\right)
\label{eq:bline}
\end{equation}
where the values of $N_A$ and $d$ are different for PI and PII
respectively, and the $(x,y,z)$ coordinate system is as shown in
Figure~\ref{fig:OWFA_aperture}.

	We then have the visibility given by \citep{Perley1989}
\begin{equation}
\vis = \dBdTnu \int_{UH} d\Omega_{\n}
T\left(\n,\nu \right) A\left(\dn,\nu
\right) e^{-2 \pi i \U\cdot \dn },
\label{eq:vis_eq1}
\end{equation}
where, $\dBdTnu = 2 k_B / \lambda^2$ is the conversion factor from
brightness temperature to specific intensity in the Raleigh - Jeans
limit, $T\left(\hat{\mathbfit{n}},\nu \right)$ is the brightness
temperature distribution on the sky and  $d\Omega_{\n}$ is
the elemental solid angle in the direction $\n$. 
For the $d\Omega_{\n}$ integral,  we use $UH,LH$ and $S$  to
denote the upper hemisphere, the lower hemisphere and the entire
celestial  sphere respectively. Note that the
integral here is over the visible upper hemisphere of the sky. 
  The entire analysis in the rest of
this sub-section is restricted   to a single frequency,  and in much
of the subsequent text   
we do not show $\nu$ explicitly.

For a telescope with a sufficiently large aperture, the primary beam
pattern $\pbnonu$ falls off rapidly as $\dn$ is increased. This restricts
the FoV to a small 
region of the sky. Considering the small FoV, we can adopt the FSA
whereby $\dn=\boldsymbol{\theta}$ is a 2D vector on
the plane of the sky with $ d\Omega_{\n} =
d^2\boldsymbol{\theta}$. It is now natural to interpret $\visnonu$
(eq.~\ref{eq:vis_eq1}) as the Fourier transform of the product
$T\left(\boldsymbol{\theta} \right) A\left(\boldsymbol{\theta}
\right)$. Here, the visibility can be expressed as a convolution \citep{Ali2014}

\begin{equation}
\visnonu = \dBdTnonu \int d^2 \U^{'} \, \apnonu \, \tilde{T}(\U^{'}) \, ,
\label{eq:vis_eq1a}
\end{equation}
where $\tilde{T}(\U^{'})$ is the Fourier transform of the
brightness temperature distribution on the sky.  We may interpret
$\visnonu$ as the weighted sum of different Fourier modes
$\tilde{T}(\U^{'})$ with the weights being given by 
the shifted aperture power pattern $\apnonu$. The  contribution peaks at $\U^{'}
= \U$ where the value of $\apnonu$ is maximum.  This tells us that we may
associate each visibility $\visnonu$ with a particular Fourier component
$\tilde{T}(\U)$ of the signal. To be more precise, the
sum extends over a width $ \vert \U -\U^{'} \vert$ which
is of the order of the dimensions of the antenna aperture in units of
the observing wavelength. As discussed earlier,
$\apnonu$ is zero beyond
the extent of the antenna aperture.

The FSA breaks down for the antennas with a small aperture which have
a large FoV. Here, one has to consider the spherical nature of the sky.
The SH function
$Y_{\ell}^m(\hat{\mathbfit{n}})$ provide a natural basis
for the signal on a spherical sky, and in the SH expansion we have 
\begin{equation}
T\left(\n,\nu \right) = \sum _{\ell,m}
a_{\ell}^m\left(\nu\right) \; Y_{\ell}^m(\n),
\label{spi2sph}
\end{equation}
where $\alm$ are the SH
coefficients of the brightness temperature distribution.

As discussed earlier by \cite{Shaw2014, Shaw2015} \citep[and also
  by][]{Zheng2014, Zhang2016a, Zhang2016b, Liu2016}, the visibilities
can be expressed in terms of the SH coefficients
$\alm$ as,
\begin{equation}
\visnonu = \sum_{\ell,m} \, \alm \,\Blmnonu .
\label{eq:vis_sph_1}
\end{equation}
The visibility expressed in eq. (\ref{eq:vis_sph_1}) is a weighted sum
of the SH coefficients $\alm$  with the weights being given by 
the beam transfer function
$\Blmnonu$. Note that this is in exact analogy with
  eq. (\ref{eq:vis_eq1a}) with the difference that in the FSA we use Fourier
  modes $\tilde{T}(\U)$ instead of the SH
  coefficients $\alm$, and we have the shifted aperture
  power pattern $\apnonu$ as the weights
  instead of $\Blmnonu$.  This analogy between
  eqs.~(\ref{eq:vis_sph_1}) and (\ref{eq:vis_eq1a}) leads to a
  picture where we interpret $\Blmnonu$ as the
  spherical sky generalization of the shifted aperture power pattern 
 $\apnonu$.

The beam transfer function $\Blmnonu$ which quantifies
how a visibility $\visnonu$ responds to a particular SH
coefficient $\alm$ can be calculated \citep{Shaw2014}
using 
\begin{equation}
\Blmnonu = \dBdTnonu \int_{UH} d\Omega_{\n} \Ylm \pbnonu e^{-2 \pi i \U\cdot \Delta
  \n} \, .
\label{eq:blm1a}
\end{equation}
 The integrand here  contains two highly oscillatory
 functions ($\Ylm$ and $ e^{-2 \pi i \U\cdot \Delta  \n}$), 
and the integral extends  over the  visible upper 
 hemisphere of the sky. It is computationally expensive and cumbersome to
 evaluate such integrals, particularly for large values of $\ell$ and
 $\U$. The major contribution to the integral comes 
 from the main lobe of the primary  beam pattern, and one may consider
using this to  restrict the integral. However the main lobe itself extends
over a large region of the sky  for telescopes with a small aperture.    
Further, the sidelobes also contribute to the integral, and it is
necessary to include the whole hemisphere to take this into account.
Finally, we note that eq.~(\ref{eq:blm1}) provides very little insight
into the behaviour of $\Blmnonu$ {\it i.e.} at which $\ell,m$ values
we have the maximum contribution for a particular baseline $\U$.

In this paper we develop a different formalism for calculating the
beam transfer function $\Blmnonu$. As noted earlier, it is
natural to interpret $\Blm$ as the spherical sky counterpart of the
shifted aperture power pattern $\apnonu$,
and in this paper we express $\Blmnonu$ as an integral of the aperture power
pattern $\apnonu$.  The $d\Omega_{\n}$ integrals in
eqs.~(\ref{eq:vis_eq1}) and (\ref{eq:blm1}) are limited  
to the visible upper hemisphere $(UH)$. The SH functions cease
to be orthonormal  when the domain is restricted to a hemisphere. The
analysis  is considerably   simplified if we assume that the primary
beam pattern and also the sky signal  
are both replicated on the lower hemisphere $(LH)$ by a reflection
with respect to the plane of the antenna aperture  (the $xy$ plane in 
Figure \ref{fig:OWFA_aperture}). The domain of
the $d\Omega_{\n}$ integrals are now extended to the   
entire sphere $(S)$. 
In this situation, the upper and lower hemispheres
both make an equal contribution to the visibility signal. We account
for this by introducing a factor of $1/2$ in the beam transfer
function which is now defined as 
\begin{equation}
\Blmnonu = \frac{\dBdTnonu}{2} \int_{S} d\Omega_{\n} \Ylm \pbnonu
e^{-2 \pi i \U\cdot \Delta   \n} \, . 
\label{eq:blm1}
\end{equation}
The fact that  $\Ylm \rightarrow (-1)^{\ell+m} \Ylm$ under
a  mirror reflection with respect to the $xy$ plane implies that
$\Blmnonu=0$ when $\ell+m$ is odd, and $\Blmnonu$ is non-zero only
when $\ell+m$ is even.  

It is necessary to note that $\alm$ in eq.~(\ref{eq:vis_sph_1}) now
refers to the SH expansion of the replicated
brightness temperature signal $T(\n)$ {\it i.e.} the signal in the
$LH$ is a mirror reflection of the signal in the visible $UH$, the 
reflection here is  with respect to the plane of the antenna aperture.

We proceed by expressing $\pbnonu$ in
eq.~(\ref{eq:blm1}) in terms of the aperture power pattern
$\apnonu$ using eq.~(\ref{eq:ft_aperture})
which gives us 
\begin{equation}
\BlmRnonu = \frac{\dBdTnonu}{2} \int_S d\Omega_{\n} \Ylm e^{-2 \pi i \U\cdot \Delta
  \n } \int d^2 \U^{'} \; e^{2 \pi i \U^{'}\cdot \Delta
  \hat{\mathbfit{n}} }\; \apnonu \, . 
\label{eq:vis_sph_2}
\end{equation}
Note that the $d\Omega_{\n}$ integral in eq.~(\ref{eq:vis_sph_2})
is over the entire sphere.

The subsequent analysis is considerably simplified if we assume that
the baselines $\U$ are coplanar with the antenna aperture. This
assumption is true for OWFA (Figure \ref{fig:aperture}). We also
expect this to be a good approximation for any compact array of fixed
antennas which point vertically overhead. Under this assumption,
$\U$ and $\U^{'}$ in eq.~(\ref{eq:vis_sph_2}) are co-planer vectors, and
we can write 
\begin{equation}
\BlmRnonu =\frac{\dBdTnonu}{2} \int d^2 \U^{'} \; \apnonu \left[ \int_S
  d\Omega_{\n} \; 
  \Ylm e^{2 \pi i \U^{'}\cdot \Delta \n }
  \right].
\label{eq:blm1a}
\end{equation}

We use the identity 
\begin{equation}
e^{2 \pi i \U^{'}\cdot \n} = 4\pi \sum_{\ell,m}
i^{\ell} \; Y_{\ell}^m (\hat{\mathbfit{U}}^{'}) \; Y_{\ell}^{m \ast}
(\n) \; j_{\ell}( 2 \pi |\U^{'}| )
\label{eq:exp2sph}
\end{equation}
to evaluate the integral in the square brackets of
eq.~(\ref{eq:blm1a}), here $j_{\ell}( 2 \pi |\U^{'}| )$ is
the $\ell^{th}$ order 
spherical Bessel function of first kind. We finally obtain 
\begin{equation}
\BlmRnonu = 2 \pi i^{\ell} \, \dBdTnonu \int d^2 \U^{'} \;  \apnonu
Y_{\ell}^m (\hat{\mathbfit{U}}^{'}) j_{\ell}( 2 \pi
|\U^{'}| )\; e^{- 2 \pi i \U^{'}\cdot \m} \, ,
\label{eq:blm2}
\end{equation}
which we use to compute  the beam transfer function $\BlmRnonu$. 

 The integrand here  contains three highly oscillatory
 functions ($Y_{\ell}^m (\hat{\mathbfit{U}}^{'})$,
 $j_{\ell}( 2 \pi |\U^{'}| )$ and $e^{- 2 \pi i \U^{'}\cdot
   \m}$), however in contrast to eq.~(\ref{eq:blm1}), the domain of the
 integral is restricted by $\apnonu$ which has compact 
support. As mentioned earlier,  $\apnonu$
is zero if $\U -\U^{'}$ exceeds the aperture dimensions in units of the
observing wavelength. This significantly reduces the computation 
for calculating $\BlmRnonu$, particularly for small apertures where the
primary beam pattern covers a large region of the sky.

\begin{figure*}
\centering
\psfrag{AA}{$\apnonu$}
\psfrag{phi}{$\phi$}
\psfrag{uxx}{$U_x$}
\psfrag{uyy}{$U_y$}
\psfrag{uuup}{$\mathbfit{U}'$}
\psfrag{uuu}{$\mathbfit{U}$}
\psfrag{bb}{$2\,b$}
\psfrag{dd}{$2\,d$}
\includegraphics[scale=0.4]{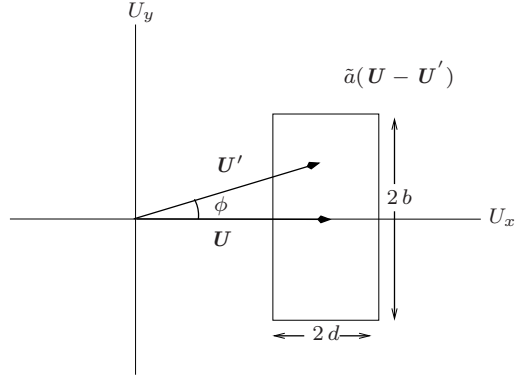}
\caption{This shows the domain of integration in eq.~(\ref{eq:blm4}). 
Here $\U$ corresponds to the baseline, $\U^{'}$ and $\phi$ can vary with 
in the aperture of the antenna (shown by the rectangle). $\phi$ becomes 
significantly small for large baseline $|\U|>>d/\lambda$. The $U_x - U_y$ 
plane here corresponds to the $\theta = \pi / 2$ for the previously adopted 
spherical polar coordinate system.} 
\label{fig:base_plane}
\end{figure*}

The subsequent analysis is restricted to the situation where the 
pointing direction  $\m$ of the antenna beam pattern  is perpendicular
to the antenna aperture so that we have $e^{- 2 \pi i \U^{'}\cdot  \m}=1$
for the phase  factor in eq.~(\ref{eq:blm2}). We adopt the coordinate
system shown in Figure \ref{fig:OWFA_aperture} with the
baseline $\U$ (Figure \ref{fig:aperture}) aligned with $x$
axis. Expressing eq.~(\ref{eq:blm2}) in  spherical polar coordinates,
the vector $\U^{'}$ is restricted  to the $\theta = \frac{\pi}{2}$ 
 plane and we then have $Y_{\ell}^m(\hat{\mathbfit{U}}^{'}) =
Y_{\ell}^m(\pi/2,0) \, e^{i m \phi}$.  
So the beam transfer function $\BlmRnonu$ can be written as 
\begin{equation}
\BlmRnonu = 2 \pi i^{\ell}\, Y_{\ell}^m (\frac{\pi}{2},0)
\, \dBdTnonu \int dU^{'} \, U^{'}\, j_{\ell}( 2 \pi 
U^{'} ) \, \int d\phi \, \tilde{a}\left(U- U^{'} \cos \phi,
U^{'} \sin \phi \right) e^{i m \phi} \, , 
\label{eq:blm4}
\end{equation}
where the domain of integration is shown in Figure
\ref{fig:base_plane}. 
The integral here contains two highly oscillatory
functions $j_{\ell}( 2 \pi U^{'} )$ and $e^{i m \phi}$. 

Considering the complex conjugate $[\BlmRnonu]^{*}$, the fact that 
$Y^{m*}_l(\theta,\phi)=(-1)^m \, Y^{-m}_l(\theta,\phi)$ implies that 
$[\BlmRnonu]^{*}={B_{\ell}^{-m}\left(\mathbfit{U} \right)}$. 
 If we further assume that the aperture
power pattern is symmetric with respect to $U_y$ {\it i.e.} 
$\tilde{a}(U_x,-U_y)=\tilde{a}(U_x,U_y)$ (which is true for OWFA) we
have $[\BlmRnonu]^{*}=(-1)^{\ell} \BlmRnonu$. This implies that 
for even $\ell$ we have the non-zero elements 
$B_{\ell}^{0}(\U),B_{\ell}^{2}(\U)=B_{\ell}^{-2}(\U),
B_{\ell}^{4}(\U)=B_{\ell}^{-4}(\U),...,B_{\ell}^{\ell}(\U)=B_{\ell}^{-\ell}(\U)$
which are all real.  
For odd $\ell$ we have the non-zero elements 
$B_{\ell}^{1}(\U)=-B_{\ell}^{-1}(\U),
B_{\ell}^{3}(\U)=-B_{\ell}^{-3}(\U),...,B_{\ell}^{\ell-1}(\U)=-B_{\ell}^{-\ell+1}(\U)$  
which are all imaginary.  The other elements of
$B_{\ell}^{m}(\U)$ are all zero.

It is possible to obtain further insight into the behaviour of 
$\BlmRnonu$ if we adopt the LA \citep{Limber1954}
\begin{equation}
j_{\ell} ( 2 \pi U^{'} ) \approx \sqrt{\frac{\pi}{2 \ell
    +1}} \delta_D \left( \ell + 1/2 -  2 \pi U^{'} \right). 
\label{eq:limber}
\end{equation}
which holds for large $\ell$. Here we also assume that the baseline is
large compared to the antenna aperture dimensions $U \gg b/\lambda$ so
that the $\phi$ range subtended by the aperture
(Figure~\ref{fig:base_plane}) is small whereby $\cos \phi \approx 1$
and $\sin \phi \approx \phi$.  We then have 
\begin{equation}
\BlmRnonu =  i^{\ell}\, \sqrt{\frac{2 \ell+1}{16 \pi}}
Y_{\ell}^m (\frac{\pi}{2},0)
\, \dBdTnonu \int d\phi \, \tilde{a}\left(U- \frac{2 \ell+1}{4 \pi}, 
\frac{2 \ell+1}{4 \pi} \phi \right) e^{i m \phi} \, .  
\label{eq:blm4b}
\end{equation}
For OWFA we can decompose the primary beam pattern (eq.~\ref{eq:pb})  as 
$A\left(\Delta \hat{\mathbfit{n}}\right)=A_x(\Delta n_x) A_y(\Delta
n_y)$ and the aperture power pattern (eq.~\ref{eq:OWFA_ap}) as 
$ \tilde{a}\left(\U^{'}\right)=\tilde{a}_x(U_x^{'}) 
\tilde{a}_y(U^{'}_y)$, where $A_x(\Delta n_x)$ and $A_y(\Delta
n_y)$ are respectively the Fourier transforms
(eq.~\ref{eq:ft_aperture})  of $\tilde{a}_x(U_x^{'})$ and  
$\tilde{a}_y(U^{'}_y)$. It is then possible to analytically evaluate
the $\phi$ integral in eq.~(\ref{eq:blm4b}) and we have 
\begin{equation}
\BlmRnonu = i^{\ell} \, \dBdTnonu\sqrt{\frac{\pi}{2 \ell +1}}
Y_{\ell}^m (\frac{\pi}{2},0) \tilde{a}_x\left(\U -\frac{2
  \ell +1}{4 \pi}\right)  A_y\left( \frac{m}{2 \ell + 1}\right) \, .  
\label{eq:blm5}
\end{equation}
where $\tilde{a}_x(U_x^{'}) =(\lambda/d)
 \Lambda(U_x^{'} \lambda / d)$ and  $A_y(\Delta n_y)={\rm sinc}^2(\pi b
 \Delta n_y/\lambda)$ for OWFA.

It is interesting to note that, in contrast to  eq.~(\ref{eq:blm1}), the
alternate expression (eq.~\ref{eq:blm2}) obtained here provides useful 
insight into the behaviour of the beam transfer function $\BlmRnonu$. We
see that the integrand in eq.~(\ref{eq:blm2}) is  the product of
$\apnonu$, which is peaked at $\U=\U^{'}$,
and $j_{\ell}( 2 \pi |\U^{'}| )$, which is peaked at $\ell= 2
\pi|\U^{'}|$. Based on this we can infer that $\BlmRnonu$ has a maximum value
when $\ell= 2 \pi|\U|$, and the value of $\BlmRnonu$ falls off as the
difference between $\ell$ and  $2 \pi|\U|$ is increased. 
In other words, the visibility $\mathcal{V}(\U,\nu)$ receives maximum
contribution from the SH coefficients $ \, \alm$ 
with $\ell=2 \pi |\U|$. The spread $\Delta \ell$ around
this $\ell$ value depends on the aperture power pattern 
with $\Delta \ell \sim 2 \pi D/\lambda$ where $D$ represents the
aperture size, {\it i.e.} we have a smaller spread in $\ell$ if the 
aperture is small as compared to a large aperture.  This behaviour 
is explicit in the  approximate analytical expression (eq. \ref{eq:blm5}) 
for $\Blmnonu$ .


\begin{figure*}
\psfrag{modflm}{$|\Flm|$}
\psfrag{mm}{$m$}
\psfrag{leq500}{$\ell = 500$}
\psfrag{leq1000}{$\ell = 1000$}
\psfrag{leq3000}{$\ell = 3000$}

\centering
\includegraphics[scale=0.6,angle=0]{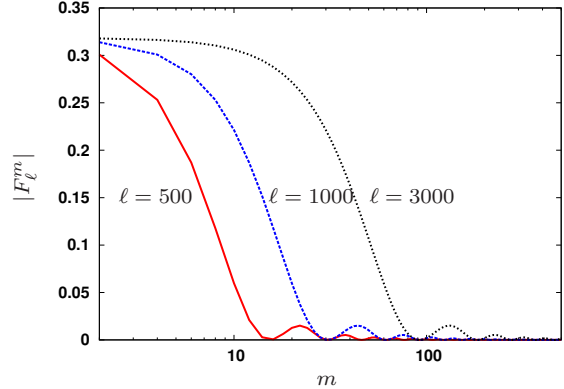}
\caption{This shows the variation of $|\Flm|$ as a function of $m$ for
  the fixed $\ell$ values $500({\rm solid \, line}), 1000({\rm dased
    \, line})$ and $3000({\rm dotted \, line})$. }
\label{fig:flm}
\end{figure*}

Eq.~(\ref{eq:blm5}) provides further insight in the behaviour of
$\Blmnonu$. For a fixed $\ell$, it is possible to analyze 
the $m$ dependence of $\Blmnonu$ by considering the function 
$\Flm = Y_{\ell}^m (\frac{\pi}{2},0) A_y\left(\frac{m}{2 \ell + 1}\right)$
which appears in  eq.~(\ref{eq:blm5}). 
Figure~\ref{fig:flm} shows the variation
of $|\Flm|$ as a function of $m$ for fixed $\ell$ values (mentioned in
the figure).  We find that the $m$ dependence of $|\Flm|$  is very similar to
${\rm sinc}^2(\pi b m/(\lambda(2 \ell+1)))$. We expect the amplitude
of $\Blmnonu$ to   be maximum for $m=0$ and decrease with 
increasing $m$. 

\begin{figure*}
\psfrag{sumwlmmm}{$\,\,\,\, |B_{\ell}^m|$}
\psfrag{l}{$\ell$}
\psfrag{mmmm0}{$m = 0$}
\psfrag{mmmm4}{$m = 4$}
\psfrag{mmmm14}{$m = 14$}
\psfrag{mmmm12}{$m = 12$}
\psfrag{bl5PI}{ $a = 5$} 
\psfrag{bl21PI}{ $a = 21$}
\psfrag{bl30PII}{ $a = 30$}
\psfrag{bl60PII}{ $a = 60$} 
\psfrag{OWFAPI}{PI}
\psfrag{OWFAPII}{PII}

\centering
\begin{minipage}{190mm}
\subfigure
    {\includegraphics[scale=0.6,angle=0]{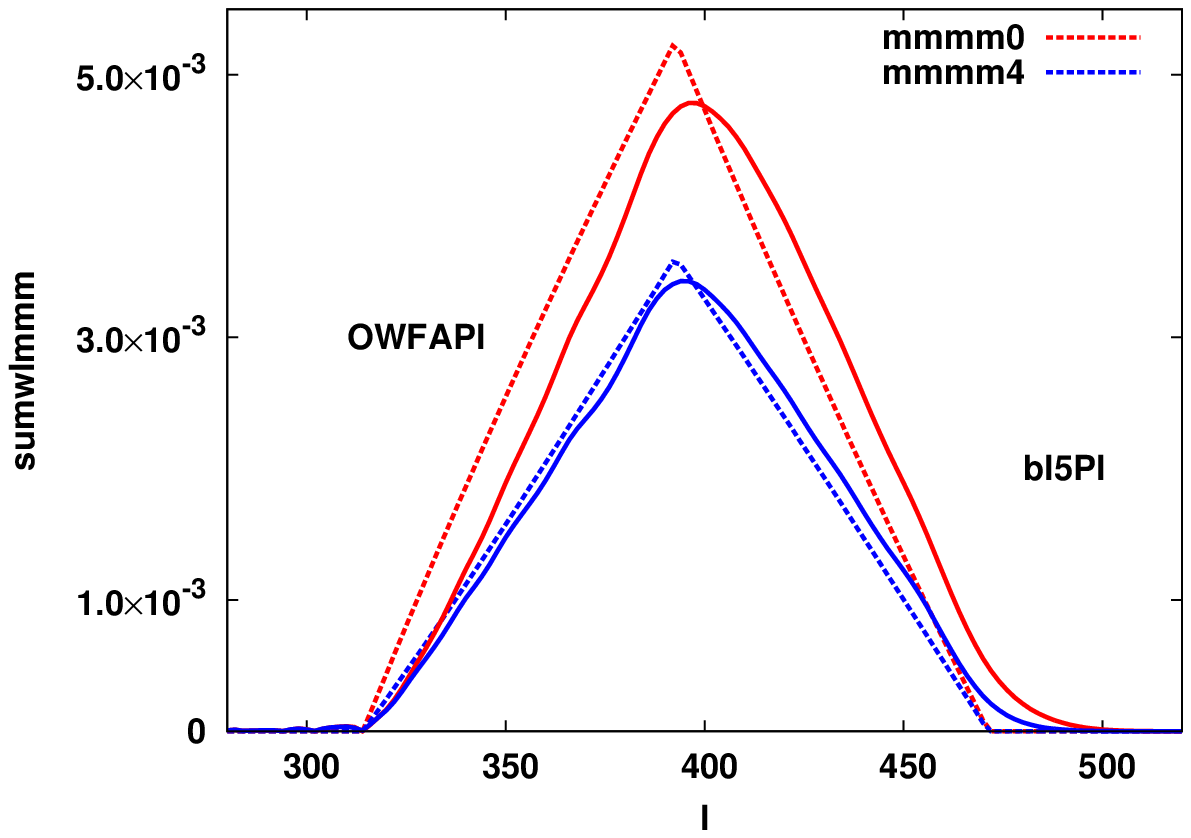}}
\hskip 7mm \subfigure
    {\includegraphics[scale=0.6,angle=0]{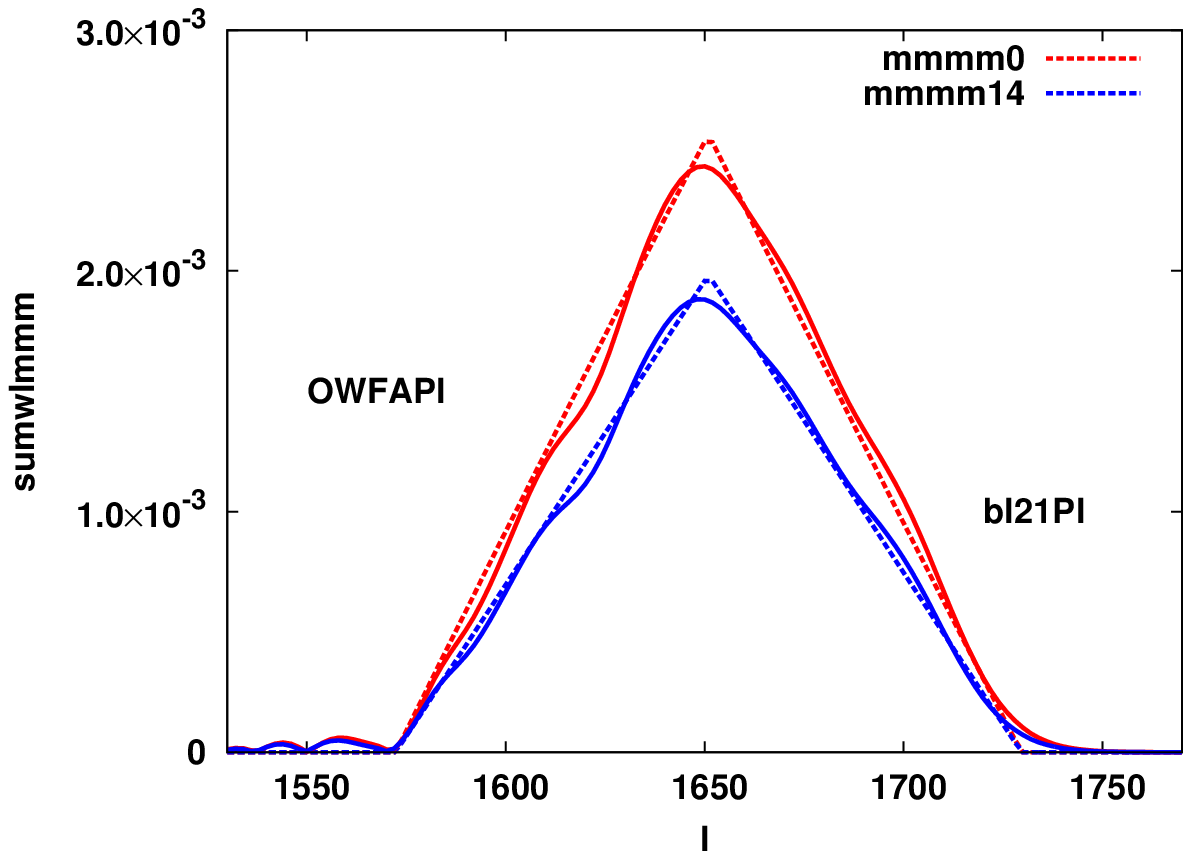}}
\end{minipage}

\begin{minipage}{190mm}
\subfigure
  {\includegraphics[scale=0.6,angle=0]{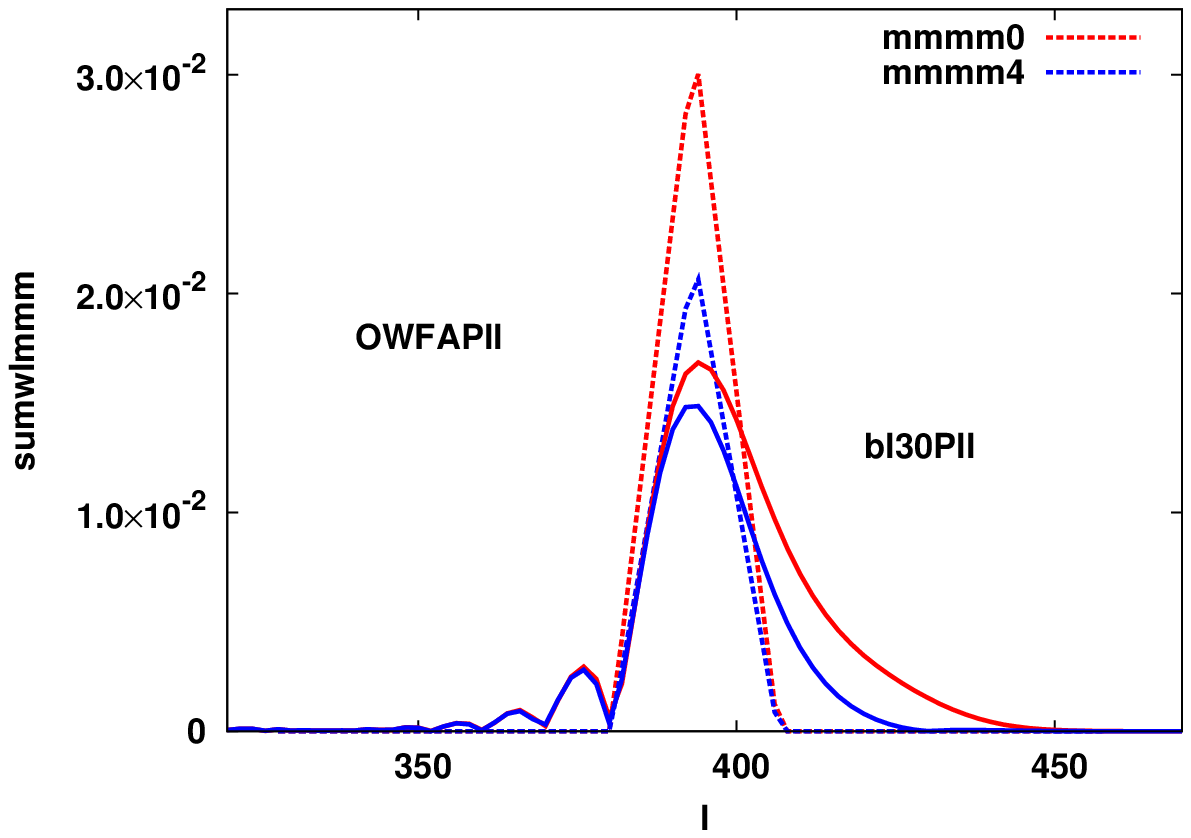}}
\hskip 7mm 
\subfigure
  {\includegraphics[scale=0.6,angle=0]{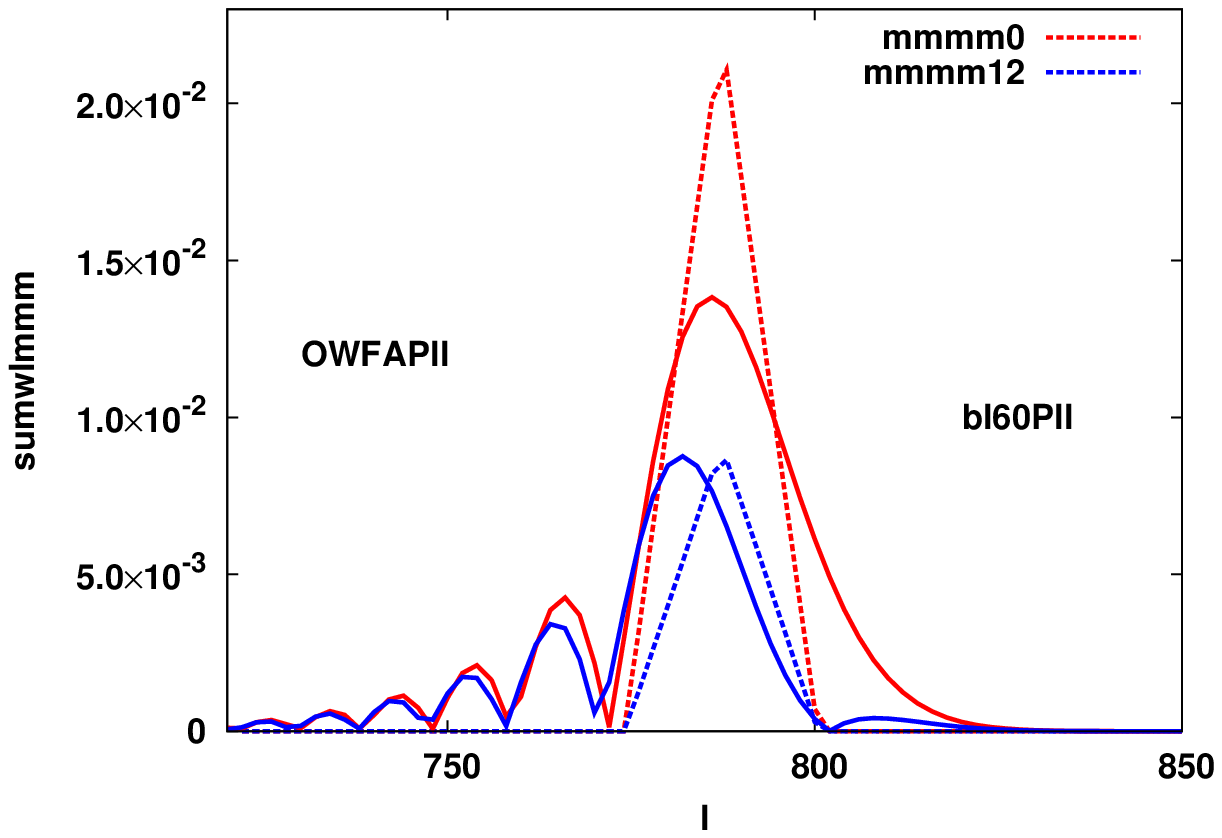}}
\end{minipage}
\caption{This shows the beam transfer function $\Blma$ as a function 
of $\ell$  for  fixed values of $\U_a$
with the $a$ and $m$ values as mentioned in the figure.
The top and bottom rows show results for 
OWFA PI and PII respectively. 
The solid and dashed lines represent $\Blma$ 
computed using eq.~(\ref{eq:blm2}) and the LA
(eq.~\ref{eq:blm5}) respectively.}
\label{fig:bt_func}
\end{figure*}

Figure~\ref{fig:bt_func} shows the OWFA beam transfer function
$\Blma$ as a function of $\ell$ for a few fixed values of $\U_a$
with the $a$ and $m$ values as mentioned in the figure. 
The solid and dashed lines  correspond to  $\Blma$  computed using 
eq.~(\ref{eq:blm2}) and the LA  (eq.~\ref{eq:blm5})
respectively.
The top row  shows the results for OWFA PI where we see that $|\Blma|$
peaks at $\ell_a =395$ and $1650$ for $a = 5$ and $21$
respectively. This  is consistent with  $\ell_a \approx 2 \pi a d /
\lambda$ where $d = 11.5 \, {\rm m}$ and $2 \pi d / \lambda \approx 79$.
We also see that, as expected, for all values of $a$ the beam 
transfer function extends over a nearly fixed
interval of $\Delta \ell = \pm 79$ around the peak  value. 
We find that LA works very well for the larger 
baseline whereas it is in reasonable agreement for the smaller baseline. 
The bottom row of Figure~\ref{fig:bt_func} 
shows the results for OWFA PII where we see that $|\Blma|$
peaks at $\ell_a =390$ and $782$ for $a = 30$ and $60$
respectively,  which is consistent with  $\ell_a \approx 2 \pi a d /
\lambda$ where $d = 1.92 \, {\rm m}$ and $2 \pi d / \lambda \approx 13$.
However, unlike  PI, here we see that $|\Blma|$  shows several secondary 
peaks (oscillations) after the
first zero to the left of $\ell_a$ and it shows a tail which extends
to $\Delta \ell \approx 40$ to the right of $\ell_a$. These features are
not present in the $\Blma$ computed using LA
and we find that the LA  does not work well even 
for the large baselines ({\it e.g.} $a = 60$).  The oscillations here
are a manifestation of the  oscillations in the 
Bessel function  $j_{\ell}(2 \pi U^{'})$ in eq.~(\ref{eq:blm4}).
For PI the aperture power pattern $\apnonu$ is wider than
the oscillation  period of  $j_{\ell}(2 \pi U^{'})$ and these
oscillations are averaged out  in eq.~(\ref{eq:blm4}). However,
PII has a smaller aperture and consequently these oscillations persist
in $\Blma$. 

\section{Visibility Correlation and Angular Power Spectrum} \label{sec:vis_cor}
The interest here is in a statistical detection of the redshifted
cosmological 21-cm signal using radio-interferometric observations.
In  this section we consider a more general situation where we wish to
detect an arbitrary statistical sky signal using radio-interferometric
observations. We assume that the brightness temperature distribution
$T\left(\n,\nu \right)$
on the sky is the outcome  of a statistically homogeneous and
isotropic random process. Note that, this assumption requires 
$T\left(\n,\nu \right)$ to be defined on the entire celestial sphere
$S$. The two point  statistics of this signal
can be completely quantified using 
the multi-frequency angular power spectrum
(MAPS; \citealt{Datta2007}) which
jointly characterises the angular and the frequency dependence
of this sky signal. The MAPS of the brightness temperature distribution at two
different frequencies $\nu_i$ and $\nu_j$ is defined through 
\begin{equation}
\maps =\left\langle a_{\ell}^{m} \left( \nu_i \right) \, a_{\ell}^{m \, *}
\left( \nu_j \right) \right\rangle .
\label{eq:maps_1}
\end{equation}
The measured visibilities $\vis$ arising from this random signal
are also random. We use two-visibility correlations
\citep{Bharadwaj2001a} to observationally quantify the statistical
properties of the input sky signal. The two-visibility correlation
\begin{equation}
\viscor \equiv \langle \mathcal{V}(\U_a,\nu_i) \, \mathcal{V}^*(\U_b,\nu_j) \rangle 
\label{eq:viscor_1}
\end{equation}   
refers to the correlation in the visibilities measured at two 
baselines $\U_a$ and   $\U_b$, and  frequencies $\nu_i$ and $\nu_j$.
The angular brackets here denote an ensemble average over different random
realization of the sky signal.  The two point statistics, or
equivalently the power spectrum (MAPS), of the sky signal is
contained in this two-visibility correlation. 

The calculations are fairly simplified in the FSA \citep{Ali2014} 
where we have 
 \begin{equation}
 \viscor = \dBdT^2 \int d^2\U^{'} \tilde{a}\left(\U_a -\U^{'}, \nu_i \right)
 \tilde{a}\left(\U_b -\U^{'}, \nu_j \right) \mathcal{C}_{2\pi U'}\left( \nu_i, \nu_j \right)\,,
 \label{eq:viscor_2}
\end{equation}  
 which relates the visibility correlation to  MAPS $\mathcal{C}_{2\pi U}\left( \nu_i, \nu_j \right)$. This can be obtained by using eq.~(\ref{eq:vis_eq1a})
 in eq.~(\ref{eq:viscor_1}), and finally utlizing the
 fact \citep{Datta2007} that  
\begin{equation}
\langle \tilde{T}(\U, \nu_i) \tilde{T}(\U^{'}, \nu_j) \rangle  = \delta_D^2 \left( \U - \U^{'}\right) \mathcal{C}_{2\pi U}\left( \nu_i, \nu_j \right)
\label{eq:maps_2}
\end{equation}
where $\delta_D^2 \left( \U - \U^{'}\right)$ is the $2$D
Dirac delta function. This expression for the  visibility correlation 
(eq.~\ref{eq:viscor_2}) is  adequate for telescopes with small FoV
(e.g. OWFA PI).

It is necessary to consider the SH formalism (eq.~\ref{eq:vis_sph_1})
for  telescopes with a large FoV (e.g. OWFA PII). To obtain an 
expression for the visibility correlation in the SH formalism, 
we revisit the visibilities as discussed earlier (in eq.~\ref{eq:vis_eq1}),
\begin{equation}
\visnonu = Q \int_{UH} d\Omega_{\n}
T\left(\n \right) A\left(\dn
\right) e^{-2 \pi i \U\cdot \dn } \, , \nonumber
\label{eq:vis_eq1a}
\end{equation}
where the $d\Omega_{\n}$ integral is over the visible upper hemisphere $(UH)$.
For our convenience, we introduce the visibility
\begin{equation}
\visnonup = Q \int_{LH} d\Omega_{\n}
T\left(\n \right) A\left(\dn
\right) e^{-2 \pi i \U\cdot \dn } \, ,
\label{eq:vis_eq1b}
\end{equation}
where the $d\Omega_{\n}$ integral is over the lower hemisphere $(LH)$ of the sky. 
Here we can safely assume that the major contributions of the sky signal 
to  $\visnonu$ and $\visnonup$ are statistically uncorrelated which implies 
that the $\visnonu$ and $\visnonup$ are statistically uncorrelated {\it i. e.}
\begin{equation}
\left\langle (\visnonu + \visnonup) (\visnonu + \visnonup)^{*} \right\rangle = \left\langle |\visnonu|^2 \right\rangle + \left\langle |\visnonup|^2 \right\rangle \, .
\label{eq:viscor_sph1}
\end{equation}
We also have,
\begin{equation}
\left\langle |\visnonu|^2 \right\rangle = \left\langle |\visnonup|^2 \right\rangle \, .
\label{eq:viscor_sph1a}
\end{equation}
Following the formalism discussed in the previous section, the 
total visibility can be expanded as, 
\begin{equation}
\visnonu + \visnonup = \sum_{\ell, m}  b_{\ell}^m \,\Blmnonu \, .
\label{eq:vis_sph_3}
\end{equation}
Here $b_{\ell}^m$ are the SH coefficients of the sky signal 
$T(\n)+T_R(\n)$, where $T_R(\n) $ is the $T(\n)$ replicated on the whole 
sphere $(S)$ after a reflection with respect to the aperture plane of the 
antenna. The statistical properties of $b_{\ell}^m$ can be quantified 
using the MAPS of the brightness temperature distribution $T(\n)$ 
(discussed in eq.~\ref{eq:maps_1}) as,
\begin{equation}
\left\langle b_{\ell}^m(\nu_i) \, b_{\ell^{'}}^{m^{'}*}(\nu_j) \right\rangle = 4 \, \mathcal{C}_{\ell}\left( \nu_i, \nu_j \right) \, \delta_{\ell \ell^{'}} \, \delta_{mm^{'}} \, .
\label{eq:cl_2}
\end{equation}
We then obtain the visibility correlation to be 
\begin{equation}
\langle \mathcal{V}(\U_a,\nu_i) \,
\mathcal{V}^*(\U_b,\nu_j) \rangle = \sum_{\ell} \,
\mathcal{C}_{\ell}\left( \nu_i, \nu_j \right) \,
\mathcal{W}_{\ell}\left(\U_a, \nu_i ;\U_b, \nu_j \right) \, ,
\label{eq:viscor_sph}
\end{equation}
where $\mathcal{W}_{\ell}$ is a  window function  which is defined
as
\begin{equation}
\mathcal{W}_{\ell}\left(\U_a, \nu_i ;\U_b, \nu_j \right)
= 2\, \sum_m \, B_{\ell}^m\left(\U_a, \nu_i \right) \,
B_{\ell}^{m*}\left(\U_b, \nu_j \right).
\label{eq:wgt_1}
\end{equation}

Eqs.~(\ref{eq:viscor_2}) and (\ref{eq:viscor_sph}) both describe how the two-visibility correlations 
$\viscor$, which can be measured directly from  observations,  is related to the statistics 
of the sky signal namely MAPS $\maps$. We see that in both cases the two-visibility 
correlation is a weighted sum of $\maps$. We have the weights 
$\tilde{a}\left(\U_a -\U \right)  \tilde{a}\left(\U_b -\U \right)$ in the FSA 
(eq.~\ref{eq:viscor_2}) where $\ell= 2 \pi|\U|$. First considering a situation 
where $\U_a = \U_b$, the weight $|\tilde{a}\left(\U_a -\U \right)|^2$  peaks 
at $\ell= 2 \pi|\U_a|$ and we have a contribution from the range of 
multipoles $\Delta \ell \sim \pm 2 \pi D/\lambda$, where $D$ is the dimension 
of the antenna aperture, the weight falls to zero beyond this interval. Considering 
two different baselines $\U_a \neq \U_b$, the weights are the
product of two functions, one which peaks at $\U = \U_a$ and another
peaks at $\U = \U_b$. The weights  
have non-zero values only if there is some overlap between $\tilde{a}\left(\U_a -\U \right)$ 
and $\tilde{a}\left(\U_b -\U \right)$ {\it i.e.} $|\U_a - \U_b| < 2 \pi D/\lambda$. There is 
no overlap and the visibilities are uncorrelated if $|\U_a - \U_b| > 2 \pi D/\lambda$. The 
properties of the FSA visibility correlation have been analyzed in detail in several 
earlier works \citep{Bharadwaj2001a, Bharadwaj2005}.

	Considering the SH formalism (eq.~\ref{eq:viscor_sph}), we see that, here also the 
weights are a product of two functions (eq.~\ref{eq:wgt_1}). Recollect that $\Blm$ 
peaks at $\ell= 2 \pi|\U|$ and has a width $\Delta \ell \sim \pm 2 \pi D / \lambda$ 
around this value. Considering the weight function for $\U_a = \U_b$ we see that this peaks 
at $\ell= 2 \pi|\U_a|$, or in other words, the visibility correlation $\viscora$ essentially 
responds to $\maps$ in a range of multipoles $\Delta \ell = \pm 2 \pi D/ \lambda$ 
peaked around $\ell= 2 \pi|\U_a|$. The fact  that the index $m$ does not appear in 
eq.~(\ref{eq:viscor_sph}) is a consequence of the assumption that the sky signal is
statistically homogeneous and  isotropic on the sky. Next, considering two different 
baseline $\U_a \neq \U_b$, here again we see that the expectations are qualitatively 
similar to these for the FSA {\it i.e.} the visibilities are correlated only if 
$|\U_a - \U_b| < 2 \pi D/\lambda$ and there is no correlation for larger separations. 
Although the visibility correlations are expected to be qualitatively similar in both 
the FSA and the \SHE, the predicted values are expected to differ. We have quantified 
these differences in subsequent sections of this paper.


\subsection{The OWFA Window Function $(\mathcal{W}_\ell)$} \label{subsec:wl}

In this subsection  we explicitly calculate the window function and discuss its behaviour 
for the two modes PI and PII of OWFA. We have used the OWFA aperture power pattern 
(eq. \ref{eq:OWFA_ap}) in eq. (\ref{eq:blm2}) to evaluate
$\Blm$.   Finally, we have   used eq. (\ref{eq:wgt_1}) to compute the
window function $\Wl$.  For comparison, we have also used LA
(eq.~\ref{eq:limber}) to calculate $\Blm$ and  used
this in eq. (\ref{eq:wgt_1}) to compute the corresponding  window
function $\Wl$.  Throughout the entire subsequent analysis we have
assumed that 
the  pointing direction $\m$ is towards the celestial equator {\it
  i.e.} normal to the aperture (Figure~\ref{fig:OWFA_aperture})
which results in $e^{- 2 \pi i   \U^{'}\cdot  \m} = 1$.

\begin{figure*}
\psfrag{sumwlmmm}{$\qquad\qquad W_{\ell}$}
\psfrag{l}{$\qquad\qquad\qquad\qquad\qquad\ell$}
\psfrag{nol}{$\,$}
\psfrag{bl5PI}{{ $a = 5$}} 
\psfrag{bl6PI}{{$a = 6$}} 
\psfrag{bl21PI}{{ $a = 21$}} 
\psfrag{bl22PI}{{$a = 22$}} 
\psfrag{ztfz}{{$\bf 350$}} 
\psfrag{zfzz}{{ $ \bf 400$}}
\psfrag{zffv}{{ $ \bf 450$}}
\psfrag{zfvzz}{{ $ \bf 500$}} 
\psfrag{zfvfvz}{{ $ \bf 550$}}
\psfrag{oszz}{{$\bf 1600$}} 
\psfrag{osfz}{{ $ \bf 1650$}}
\psfrag{osvzz}{{ $ \bf 1700$}}
\psfrag{osvfz}{{ $ \bf 1750$}} 
\psfrag{oezz}{{ $ \bf 1800$}}
\psfrag{zr}{$0$}
\psfrag{5powm6}{$ \bf 5\times10^{-6}$}
\psfrag{1powm5}{$ \bf 1\times10^{-5}$}
\psfrag{15powm5}{$ \bf 1.5\times10^{-5}$}
\psfrag{2powm5}{$ \bf 2\times10^{-5}$}

\psfrag{bl30PI}{{$a = 30$}}
\psfrag{bl31PI}{{$a = 31$}}
\psfrag{bl60PI}{{$a = 60$}}
\psfrag{bl61PI}{{$a = 61$}}
\psfrag{ztsz}{{$\bf 360$}} 
\psfrag{zfzz}{{ $ \bf 400$}}
\psfrag{zffz}{{ $ \bf 440$}}
\psfrag{zsfz}{{ $ \bf 750$}} 
\psfrag{zsnz}{{ $ \bf 790$}}
\psfrag{zetz}{{ $ \bf 830$}}
\psfrag{zr}{$0$}
\psfrag{m1powm4}{$ \bf -1\times10^{-4}$}
\psfrag{1powm4}{$ \bf 1\times10^{-4}$}
\psfrag{2powm4}{$ \bf 2\times10^{-4}$}
\psfrag{3powm4}{$ \bf 3\times10^{-4}$}
\psfrag{4powm4}{$ \bf 4\times10^{-4}$}
\psfrag{OWFAPI}{PI}
\psfrag{OWFAPII}{PII}

\centering
\begin{minipage}{170mm}
\subfigure[]
  {\label{fig:win_func-PI}
  \includegraphics[scale=1.2,trim= 0 50 0 50,angle=0]{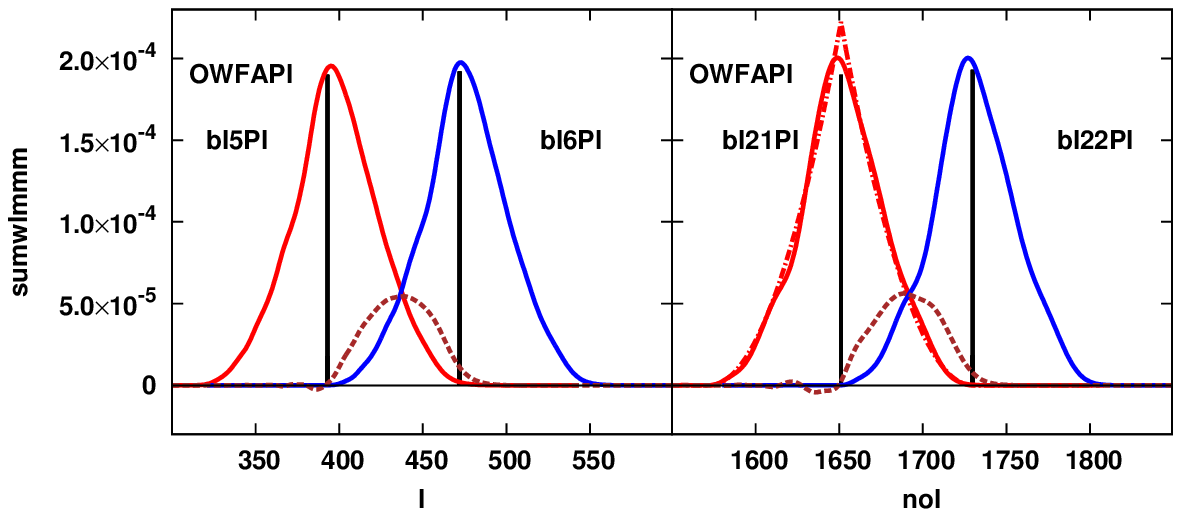}}

\subfigure[]
{\label{fig:win_func-PII}
  \includegraphics[scale=1.2,trim= 0 50 0 50,angle=0]{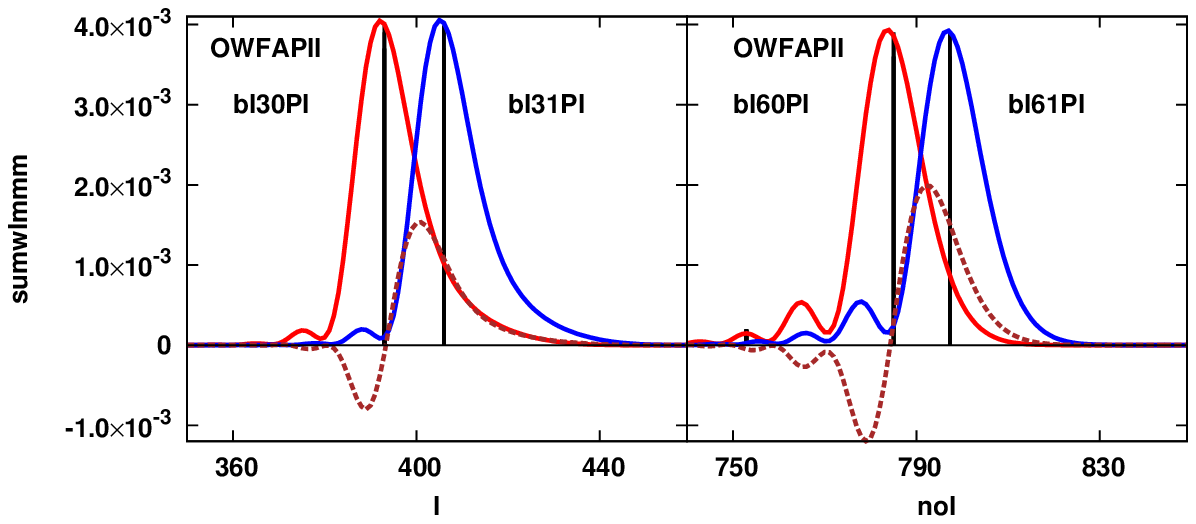}}
\end{minipage}
\caption{This shows the window functions for OWFA PI and PII. Solid 
lines in the top row show the window functions $\Wlsamec$ for the same 
baselines of OWFA PI for $a = 5,6,21$ and $22$. The dot-dashed line shows the $\Wlsamec$ 
evaluated using LA  for $a = 21$.
Solid lines in the bottom row show the window functions $\Wlsamec$ for the same 
baselines of OWFA PII for $a = 30,31,60$ 
and $61$. The dashed lines in the figure represents the window functions 
$\Wladjc$ for the corresponding adjacent baselines. }
\label{fig:win_func}
\end{figure*}

We first consider the OWFA window function $\Wl$ for a situation where
we have the same baseline $\U_a=\U_b=a d/\lambda$. Based on our
earlier discussion, the OWFA 
window function $\Wlsame$   is expected to peak at $\ell_a = 2 \pi a d
/ \lambda$ and have a width of $\Delta \ell =  \pm 2 \pi d / \lambda $
around $\ell_a$. 
Figure~\ref{fig:win_func} shows the OWFA window function
$\Wl$ as a function of $\ell$ for a few fixed values of $\U_a$ 
 (mentioned in the figure legend),  and fixed frequency 
$\nu_i = \nu_j = \nu_c = 326.5 \, {\rm MHz}$, which is
the nominal frequency for OWFA.
The top row  shows the results for OWFA PI where we see that $\Wlsamec$
peaks at $\ell_a =395,471,1650$ and $1725$ for $a = 5,6,21$ and $22$
respectively  which is consistent with  $\ell_a \approx 2 \pi a d /
\lambda$ where $d = 11.5 \, {\rm m}$  
and $2 \pi d / \lambda \approx 79$. We also see that, as expected,
for all values of $a$ the window function extends over a nearly fixed
interval of $\Delta \ell = \pm 79$ around the peak  values.
The dot-dashed line in the right panel of the top row shows the window 
function calculated using  LA for 
$a = 21$. We find that LA  works well in the large 
baseline regime.  
The bottom row  shows the results for OWFA PII where we see that $\Wlsamec$
peaks at $\ell_a =390,405,782$ and $797$ for $a = 30,31,60$ and $61$
respectively which is consistent with  $\ell_a \approx 2 \pi a d /
\lambda$ where $d = 1.92 \, {\rm m}$   and $2 \pi d / \lambda \approx 13.13$.
Considering the width of the window function, we see that for all
values of $a$, $\Wlsamec$ goes to zero at $\Delta  \ell \approx 13$ to
the  left of $\ell_a$. However, unlike  PI, here we see that
$\Wlsamec$  shows several secondary peaks (oscillations) after the
first zero to the left of $\ell_a$ and it shows a tail which extends
to $\Delta \ell \approx 40$ to the right of $\ell_a$. The oscillatory
features and the extended tail are both a consequence of the spherical
Bessel function $j_{\ell}( 2 \pi |\U^{'}| )$ in
eq.~(\ref{eq:blm2}). For PI these oscillations and the tail 
are washed out because of the integral over the broader aperture power
pattern $ \tilde{a}\left(\U_{a} - \U^{'} \right)$.

We next consider the OWFA window function $\Wl$ for a situation where
we have the two different baselines $\U_a \neq \U_b$. The aperture
power patterns $\tilde{a}\left(\U_a  - \U \right)$ and 
$\tilde{a}\left(\U_{b} - \U \right)$ have an overlap only if
$b=a \pm 1$ {\it i.e.} the overlap is restricted to the adjacent 
baselines and does not extend beyond (Figure 2 of \citealt{Ali2014}).
We further expect  the window function $\Wladjc$ for the adjacent
baselines to peak at $\ell_a = \pi (2 a \pm 1) d / \lambda$
which corresponds to the average of the two baselines  and have a 
width  $\Delta \ell =  \pm \pi d / \lambda $ which is half the width
of $\Wlsamec$. For both PI and PII we find that the peaks of the
window function  $\Wladjc$ are located at the expected $\ell$ values,
the width also is consistent with the expectations for PI. However,
for PII we find that $\Wladjc$ oscillates to the left of the peak and
has an extended tail to right. As mentioned  earlier, these are 
consequences of the spherical Bessel function $j_{\ell}( 2 \pi
|\U^{'}| )$ in eq.(\ref{eq:blm2}). We finally note that, for both
PI and PII there is an $\ell$ range where the window function
$\Wladjc$ become negative, while this is relatively prominent for PII
this feature is also present for PI.

\section{Simulation} \label{sec:simulation}
In the previous section we have developed a formalism which 
relates the two-visibility correlations $\viscor$, which can be measured 
from the observations, to the statistics of the sky signal namely MAPS $\maps$. 
In this section we present  simulations that we have carried out  to
validate this theoretical formalism. The entire analysis is restricted
to a single frequency  $\nu_c = 326.5 \, {\rm MHz}$ and do not show
this explicitly in much of the subsequent text. 

We assume that the sky signal $T(\n)$ is a Gaussian random field with an
angular power spectrum $C_{\ell}$ which we have modelled as a power law 
\begin{equation}
\mathcal{C}_{\ell} = A \, \left(\frac{\ell}{\ell_0}\right)^n \,,
\label{eq:cl_1}
\end{equation}
arbitrarily normalized  to unity ({\it i.e.} $A = 1$) at $\ell =
\ell_0=1$.  We have considered the power law index to have value 
$n=-2$ in our analysis.

The simulations were carried out using the package HEALPix
\citep[Hierarchical Equal Area isoLatitude Pixelization of a
sphere;][]{Gorski2005}. For OWFA, the largest baseline corresponds to
an angular multipole  
$\ell \approx 2 \pi \times 530 \, {\rm m}/\lambda_c=3624$. In the
simulations, we have  set $\ell_{max}= 4096$ which corresponds to  a 
pixel size of $1.712^{'}$. We have used the SYNFAST routine of
HEALPix to generate different statistically independent realization of
the sky signal $T(\n)$ corresponding to the  input $C_{\ell}$. 

\begin{figure*}
\centering \includegraphics[scale=0.35]{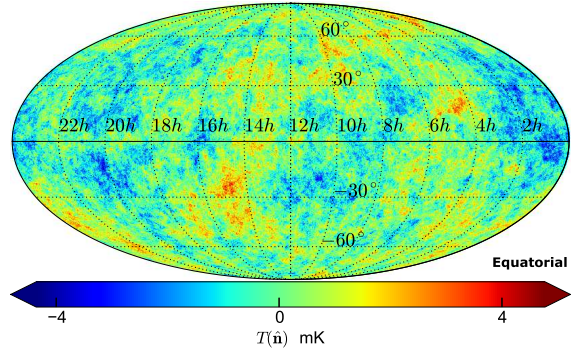}
\caption{This shows a single realization ofthe 
brightness temperature fluctuations used in the simulations.}
\label{fig:l-2}
\end{figure*}

We have computed the OWFA visibilities from the simulated  maps using 
eq.~(\ref{eq:vis_eq1}). Here, it is convenient to adopt the coordinate
system defined in Figure 2 of \citet{Marthi2016a} whereby 
\begin{equation}
\mathcal{V}(\U_a) = Q \, \Delta \Omega_{pix} \, \sum _{p=0}^{N_{pix} - 1} \, T(\alpha_p,
\delta_p) \, A(\alpha_p, \delta_p) e^{-2 \pi i U_a \, (\sin
  \delta_p - \sin \delta_0)},
  \label{eq:vis_sim1}
\end{equation}
where $\Delta \Omega_{pix}$ refers to the solid angle subtended by
each simulation pixel, $(\alpha_p, \delta_p)$ refers to the (RA, DEC) of
the $p$-th pixel, and $(\alpha_0, \delta_0)$ refers to  the pointing
direction. 
The sum here runs over all the pixels ($N_{pix}$ in number)
in the simulation.

 The OWFA primary beam pattern
$A(\Delta \n)$  eq.~(\ref{eq:pb}) can be expressed as 
\begin{equation}
A(\alpha_p, \delta_p) = {\rm sinc}^2 \left( \frac{\pi b \nu_c }{c} \,
{\rm cos} \delta_p \, {\rm sin} \left(\alpha_p - \alpha_0 \right)
\right) {\rm sinc}^2 \left( \frac{\pi d \nu_c}{c}  \left({\rm sin}
\delta_p - {\rm sin} \delta_0\right) \right). 
\label{eq:pb_sim}
\end{equation}
As mentioned earlier, we have used $(\alpha_0, \delta_0)=(0, 0)$
for the simulations presented here. 

Figure \ref{fig:l-2} shows the brightness temperature fluctuations
corresponding to a particular realization of the simulations.
We have used an ensemble of $100$ independent realizations of the
simulated sky  signal to estimate the visibility correlation
$\viscornonu$ and its variance.

\section{Results} \label{sec:results}

\begin{figure*}
\psfrag{V2V2UaUb}{$\viscornonu {\rm Jy}^2$}
\psfrag{U}{$\qquad \qquad \qquad \qquad \U$}
\psfrag{phaI}{PI} 
\psfrag{SameBaseline}{ Same Baseline}
\psfrag{AdjacentBaseline}{Adjacent Baseline}
\psfrag{1theoretical}{ Theory} 
\psfrag{1simulation}{ Simulation}
\psfrag{error}{$\, \, \, \Delta$}
\centering
 \includegraphics[scale=1.0,angle=0]{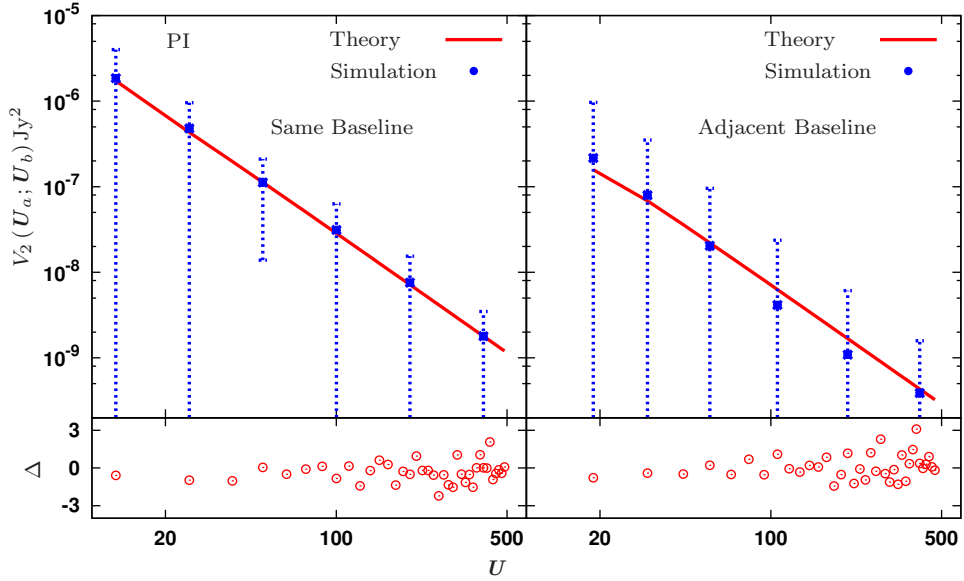}
\caption{The upper left and right panels  show the two-visibility 
correlation $\viscornonu$ for the same and the adjacent baselines respectively for 
OWFA PI. The solid lines show the theoretical predictions (eq.~\ref{eq:viscor_sph}) 
while  the  points with error-bars show the mean and  standard
deviation  from the simulations considering a few selected baselines.  
The points in the bottom panels show $\Delta$   (eq.~\ref{eq:Delta})   
which quantifies the deviation between the theoretical predictions and
the simulations. }
\label{fig:PI_APS}
\end{figure*}

The upper panels of Figure~\ref{fig:PI_APS} show the two-visibility correlation 
$\viscornonu$ for OWFA PI considering the fixed frequency $\nu_i = \nu_j = \nu_c$, 
which has not been shown explicitly. The left panel considers the situation 
where the two baselines are same $\U_a = \U_b$, whereas the right panel shows 
the results for the adjacent baselines $\U_b = \U_{a+1}$. We have used the 
$\mathcal{C}_{\ell}$ given by eq.~(\ref{eq:cl_1}) in eq.~(\ref{eq:viscor_sph}) 
to calculate the full SH theoretical predictions
for the two-visibility correlations  
$\viscornonu$ shown by the solid lines in Figure~\ref{fig:PI_APS}. We find 
that for both the same and the adjacent baselines, the theoretical predictions 
are in good agreement with the simulation. Note that $N_r = 100$ independent 
realizations were used to estimate the mean $\viscornonu$ (shown with points) 
and the standard deviation $\sigma$ (shown with error bars). The lower panels 
of Figure~\ref{fig:PI_APS} show 

\begin{equation}
\Delta = \frac{\delta \viscornonu \, \sqrt{N_r}}{\sigma} \, ,
\label{eq:Delta}
\end{equation}
where $\delta \viscornonu$ is the difference between the theoretical predictions 
and the simulations. We expect, $\Delta$ to have a Gaussian distribution with zero
mean and unit variance. We find that the values of $\Delta$ in the
lower panel of  
Figure~\ref{fig:PI_APS} are roughly centred around zero and distributed within 
$\pm 3$, consistent with what one would expect from the Gaussian
distribution. 

\begin{figure*}
\psfrag{V2V2UaUb}{$\viscornonu {\rm Jy}^2$}
\psfrag{U}{$\qquad \qquad \qquad \qquad \U$} 
\psfrag{phaII}{PII} 
\psfrag{SameBaseline}{ Same Baseline}
\psfrag{AdjacentBaseline}{ Adjacent Baseline}
\psfrag{1theoretical}{ Theory} 
\psfrag{1simulation}{ Simulation} 
\psfrag{error}{$\, \, \, \Delta $}
\centering
\includegraphics[scale=1.0,angle=0]{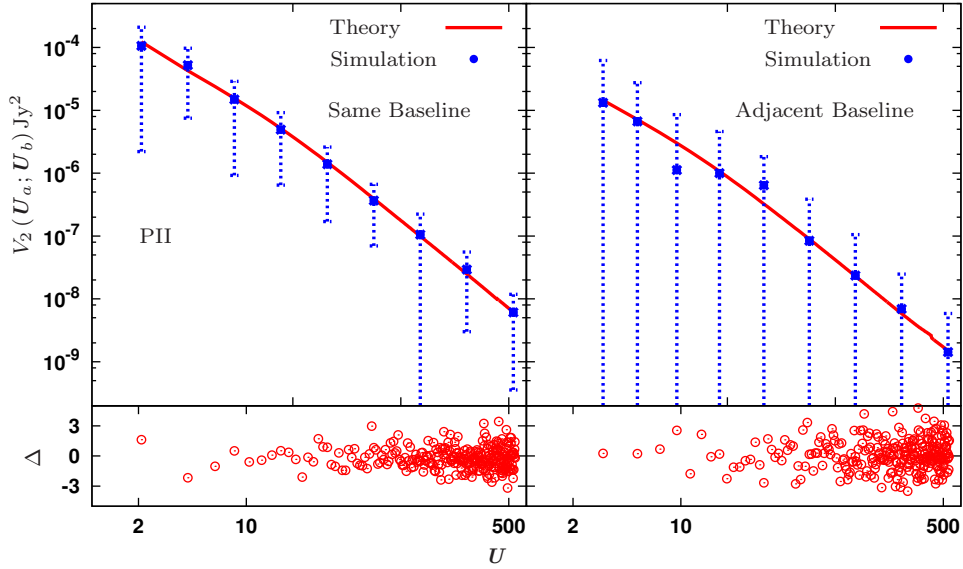}
\caption{The same as Figure~\ref{fig:PI_APS} for OWFA PII.}
\label{fig:PII_APS}
\end{figure*}

Figure~\ref{fig:PII_APS} shows a comparison of the theoretical
predictions with the simulations for the two  visibility correlations 
for OWFA PII. Here also we find that the theoretical predictions 
are in good agreement  with the simulations. The fact that the 
theoretical predictions are in good agreement with  the simulations 
for both OWFA PI and PII 
validates our formalism for calculating the beam transfer function
(eq.~\ref{eq:blm2}), the window function (eq.~\ref{eq:wgt_1}) and the
visibility correlation (eq.~\ref{eq:viscor_sph}). 
 
\begin{figure*}
\centering 
\psfrag{U}{$\qquad \qquad \qquad \qquad \U$}
\psfrag{V2V2UaUb}{$\viscornonu {\rm Jy}^2$}
\psfrag{phaI}{PI}
\psfrag{covcovcovcovcvo}{{ \qquad \qquad \, FSA}}
\psfrag{sphsphsphsphsph}{{\qquad\qquad \quad \SHE}}
\psfrag{limberapproximation}{ \qquad \qquad \quad \, LA}
\psfrag{error}{$\%$-error}
\psfrag{SameBaseline}{\, \, Same Baseline}
\psfrag{AdjacentBaseline}{Adjacent Baseline}
\includegraphics[scale=1.0,angle=0]{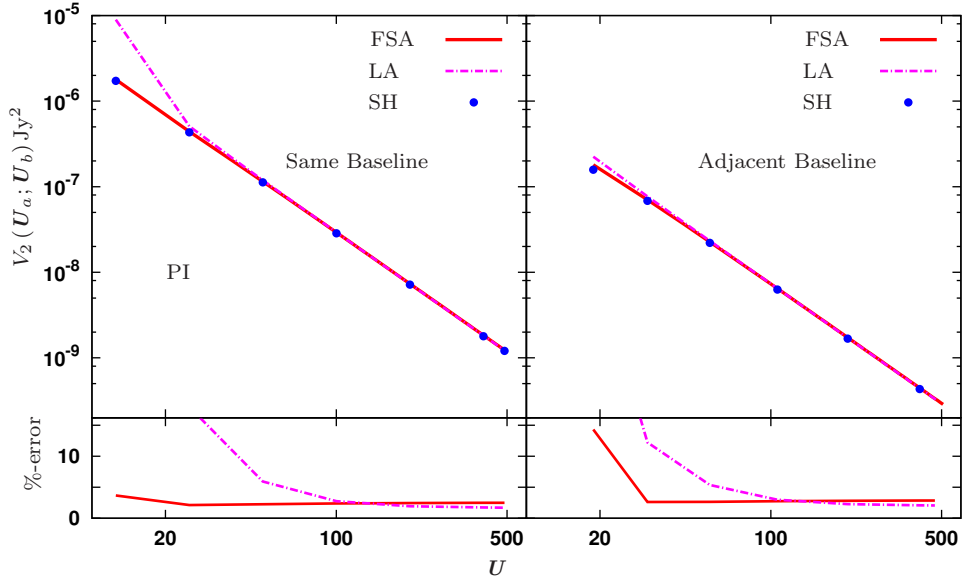}

\caption{The upper left and right panels  show the two-visibility 
correlation $\viscornonu$ for the same and the adjacent baselines respectively for
OWFA PI. The solid lines show  the theoretical predictions obtained using the FSA
(eq.~\ref{eq:viscor_2}), the points correspond to the predictions 
obtained from the full SH formalism (eqs.~\ref{eq:blm2}, \ref{eq:viscor_sph}, \ref{eq:wgt_1})and the 
dot-dashed lines show predictions obtained using the SH formalism with LA  for $\Blmnonu$ 
(eqs.~\ref{eq:blm5}, \ref{eq:viscor_sph}, \ref{eq:wgt_1}). In the lower panels, 
we show the percentage deviation of the FSA (in solid lines) and 
LA (in dot-dashed lines)predictions relative to the full SH analysis.}
\label{fig:PI_flat_APS}
\end{figure*}

Considering the two-visibility correlation $\viscornonu$  for OWFA PI, 
the upper panels of Figure~\ref{fig:PI_flat_APS} show a
comparison between  the FSA (eq.~\ref{eq:viscor_2}) (solid lines),  
the full SH formalism
(eqs.~\ref{eq:blm2}, \ref{eq:viscor_sph}, \ref{eq:wgt_1})  (points), 
 and the SH formalism with LA  for $\Blmnonu$ 
(eqs.~\ref{eq:blm5}, \ref{eq:viscor_sph}, \ref{eq:wgt_1})  
(dot-dashed lines).  The lower panels show the percentage deviation of
 the FSA and LA predictions relative to the full SH analysis. 
Considering the correlations at the same baseline (left panels), we find
that the FSA is in good agreement with the full SH predictions over
the entire $U$ range and the  deviations are less than $5 \%$. 
Considering the LA, we find that the results match  the full SH 
analysis at large baselines, the deviations are $\le 5 \%$ for $U \ge
60$. However, these deviations increase rapidly at smaller baselines
and we have $\sim 100 \%$ deviations at the smallest baseline$(U =
12.5)$. For the correlations at the adjacent baselines (right panel), we find
that the FSA is in good agreement with the full SH predictions for 
$U \ge 40$ and the deviation goes upto $14 \%$ at the smallest baseline 
$(U = 18.25)$. Considering the LA, we find that the results match  the full SH 
analysis at large baselines, the deviations are $\le 5 \%$ for $U \ge
60$. However, these deviations increase rapidly at smaller baselines
and we have $\sim 42 \%$ deviations for the correlation between  the
two smallest baselines which corresponds to a mean value of  $U =
18.25$.   

\begin{figure*}
\centering 
\psfrag{U}{$\qquad \qquad \qquad \qquad \U$} 
\psfrag{V2V2UaUb}{$\viscornonu {\rm Jy}^2$}
\psfrag{phaII}{PII}
\psfrag{covcovcovcovcvo}{{\qquad\qquad \, FSA}}
\psfrag{sphsphsphsphsph}{{\qquad \qquad \quad \SHE}}
\psfrag{limberapproximation}{ \qquad \qquad \quad \, LA}
\psfrag{error}{$\%$-error}
\psfrag{SameBaseline}{ Same Baseline}
\psfrag{AdjacentBaseline}{ Adjacent Baseline}
 \includegraphics[scale=1.0,angle=0]{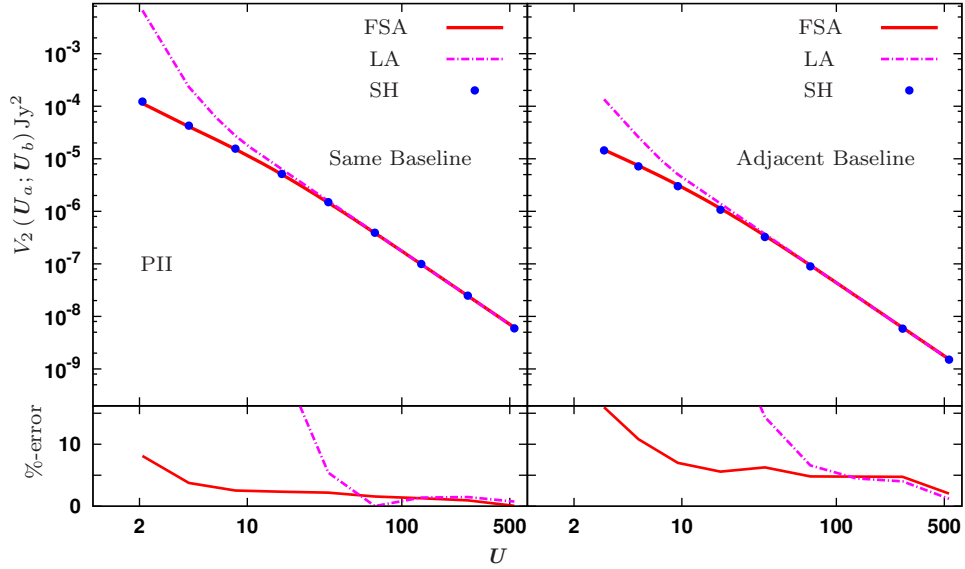}
\caption{The same as Figure~\ref{fig:PI_flat_APS} for OWFA PII.}
\label{fig:PII_flat_APS}
\end{figure*}

Figure~\ref{fig:PII_flat_APS} shows a similar comparison  for OWFA
PII. Considering the correlations at the same baseline (left panel),
we find that the FSA is in good agreement with the full SH 
predictions over the entire $U$ range and the  deviations are less
than $10 \%$.  
Considering the LA, we find that the results match  the full SH 
analysis at large baselines, the deviations are $\le 5 \%$ for $U \ge
60$. However, these deviations increase rapidly at smaller baselines
and we have more than $ 200 \%$ deviations at the smallest baseline $(U =
2.1)$. For the correlations at the adjacent baselines (right panel), we find
that the FSA is in good agreement with the full SH predictions for 
$U \ge 60$,  and the  deviation goes upto $16 \%$ at the smallest baseline pair $(U =
3.15)$. Considering the LA, we find that the results match the full SH 
analysis at large baselines, the deviations are $\le 5 \%$ for $U \ge
60$. However, these deviations increase rapidly at smaller baselines
and we have $\sim 80 \%$ deviations for the smallest baseline pair $(U =
3.15)$. 

\section{Summary and Conclusion} \label{sec:discussion}
The upcoming Ooty Wide Field Array (OWFA;
\citealt{Subrahmanya2016a}) has the detection of the redshifted 21-cm
power spectrum as one  of its primary science goals.  Earlier works
({\it e.g.} \citealt{Ali2014}) dealing with the signal and foreground
predictions for the two modes (PI and PII) of OWFA have all assumed
the FSA. However, OWFA PII has a rather
large FoV ($\sim 57^{\circ}$) in the North-South direction
and it is important to incorporate the spherical nature of the sky. In
this paper we investigate the importance of this effect relative  to
the earlier FSA. To this end we adopt the  SH analysis which relates the measured
visibilities to the SH coefficients of the sky signal
through a beam transfer function \citep{Shaw2014}. 
 
Considering a radio interferometer where the baselines are
all coplanar with the antenna aperture ({\it e.g.} OWFA, CHIME), we have 
developed a new  formalism to compute the   beam transfer
function $\Blmnonu$ as an  integral of the aperture power pattern
$\apnonu$ (eq.~\ref{eq:blm2}).  In addition to computational
advantages, our formalism provides insight into the behaviour of
$\Blmnonu$, {\it i.e.} the $\ell$ and $m$  values at which we expect this to peak
for a particular baseline $\U$.  For OWFA, 
we find that $\Blmnonu$  is expected to
peak at $\ell \approx 2 \pi U$ and fall off with increasing 
$|\Delta \ell |$  with a width $\Delta \ell \approx \pm 2
\pi d/\lambda$ around the peak value.  We also provide a closed form
analytical expression for  $\Blmnonu$  using the LA which is expected to hold at large $\ell$
or equivalently at large $U$.  This analytic expression
(eq.~\ref{eq:blm5}) indicates that $\Blmnonu$ is expected to peak at
$m=0$ and fall off with increasing $m$ with a width $\Delta m \approx
\pm (2 \ell + 1) \lambda/b$ (Figure \ref{fig:flm}). It is worth noting
that while $m$ takes values in the range $-\ell \le m \le \ell$, the 
sky signal is restricted to a range $-\ell \lambda/b  \le m
\le \ell\lambda/b $ where $\lambda/b \approx 0.03$ for OWFA {\it i.e.}
OWFA only responds to  $3 \%$ of the available $m$-modes because of
the relatively  large width $b$ of the telescope aperture. 
Figure \ref{fig:bt_func} shows $\Blmnonu$ for a few of the
baselines. For both  PI and PII we find that the $\ell$ value where 
$\Blmnonu$ peaks is roughly consistent with the expected values, for PI
the width also is consistent with this. However for PII  $\Blmnonu$
oscillates to the left of the peak and has an extended tail to the
right of the peak.  The LA provides a reasonable
match for PI, particularly at large $U$. For PII, LA  correctly
predicts the peak,  however, it does not provide a
good match to the behaviour  of  $\Blmnonu$ away from the peak  even at
large $U$.  

In the FSA, the two-visibility correlation
provides a direct estimate of  the redshifted 21-cm power spectrum 
\citep{Bharadwaj2001a,Bharadwaj2005}. In this paper we 
incorporate the  spherical sky and express the two-visibility 
correlation (eq.~\ref{eq:viscor_sph}) as a weighted sum of  
the multi-frequency angular power spectrum
(MAPS; \citealt{Datta2007}).  The weights here are  given by the window 
functions $\mathcal{W}_{\ell}\left(\U_a, \nu_i ;\U_b, \nu_j \right)$
which are a sum (over $m$) of  products of the beam transfer function  
(eq.~\ref{eq:wgt_1}). The signal,
for OWFA,  is present in the correlations at the same baseline and the
adjacent baselines only \citep{Ali2014}.  Figure \ref{fig:win_func}
shows the window function for a few select baselines. We see that the
correlation at baseline $U_a$ pick up the signal corresponding to
$\ell \approx 2 \pi U_a$ whereas the correlation between the adjacent
baselines $U_a$ and $U_b$ corresponds to $\ell \approx \pi
(U_a+U_b)$. The LA provides a good match to the window function at
large baselines for PI, however the match (not shown in the Figure) 
is not very good for PII.

We have carried out simulations to validate our spherical sky
formalism. We find (Figures \ref{fig:PI_APS} and \ref{fig:PII_APS})
that for both PI and PII our analytical  predictions are consistent
with the results from the  simulations, thereby validating our
formalism. Figures \ref{fig:PI_flat_APS} and \ref{fig:PII_flat_APS}
show a comparison between the predictions of our SH analysis with
those from the FSA. The situation where LA has been used to calculate
the beam transfer function is also considered for comparison. We find
that for both PI and PII, the FSA matches the SH predictions to within 
$16 \%$ across the entire baseline range. As expected, the match is
better for PI as compared to PII. The match is also better for the
correlations at the same baseline as compared to the adjacent
baselines. In contrast, the LA provides a good match $< 5 \%$ only at
the large baselines $U \ge 60$. In conclusion, we note that the flat
sky approximation  matches the full SH analysis to
within $15-16 \%$. It is necessary to use the latter if an accuracy
higher than this is required. 

Finally, we note that the entire analysis here is restricted to
observations in a single FoV. This has  advantage in
allowing detailed foreground modelling of the particular field which
may be useful in foreground removal. In contrast, the ``$m$-mode''
analysis proposed by \citet{Shaw2014} considers drift scan
observations which cover a large fraction of the sky. The latter
increases the available  signal, this could however make detailed
foreground modelling more difficult. We plan to study $m$-mode
analysis in the context of OWFA in subsequent work.   

\section*{Acknowledgement}
The authors would like to thank Jayaram N Chengalur and Visweshwar Ram Marthi for their 
useful suggestions and discussions. SC acknowledges the University Grant Commission, India for
providing  financial support through Senior Research Fellowship. SC would also
like to thank Sukannya Bhattacharya, Abinash Kumar Shaw, Anjan Kumar Sarkar and Debanjan Sarkar for 
their help and useful discussions. 

\bibliographystyle{mnras} \bibliography{sph_harmonics.bbl}
\end{document}